\documentclass[%
reprint,
superscriptaddress,
amsmath,amssymb,
aps,
pre,
floatfix,
]{revtex4-2}

\usepackage[table]{xcolor}
\usepackage{graphicx}% Include figure files
\usepackage{dcolumn}% Align table columns on decimal point
\usepackage{bm}% bold math
\usepackage{subcaption} % for subfigures
\usepackage{braket}
\usepackage[justification=Justified, format=plain]{caption}
\usepackage{comment}

\begin{document}

\preprint{APS/123-QED}

%\title{General distribution for the ratio of consecutive level spacing of many-body Hamiltonians}

%\title{Ratio of consecutive level spacing distribution for crossover ensembles} 
\title{Consecutive-gap ratio distribution for crossover ensembles}
%\\

\author{Gerson C. Duarte-Filho}
%\affiliation{Instituto de Física de São Carlos, Universidade de São Paulo, 13560-970 São Carlos, São Paulo, Brazil}
%
\affiliation{Departamento de F\'isica, Universidade Federal de Sergipe, 49100-000, S\~ao Crist\'ov\~ao, SE, Brazil}
\author{Julian Siegl}
\affiliation{Institute for Theoretical Physics, University of Regensburg, Regensburg, Germany}

\author{John Schliemann}
\affiliation{Institute for Theoretical Physics, University of Regensburg, Regensburg, Germany}
\author{J. Carlos Egues}
\affiliation{Instituto de Física de São Carlos, Universidade de São Paulo, 13560-970 São Carlos, São Paulo, Brazil}
%
%\affiliation{Department of Physics, University of Basel, Klingelbergstrasse 82, CH-4056 Basel, Switzerland }

\date{\today}% It is always \today, today,

\begin{abstract}

The study of spectrum statistics, such as the consecutive-gap ratio distribution, has revealed many interesting properties of 
many-body complex systems. Here we propose a two-parameter surmise expression 
for such distribution to describe the crossover between the Gaussian orthogonal ensemble (GOE) and Poisson statistics.
This crossover is observed in the isotropic Heisenberg spin-$1/2$ chain with disordered local field, exhibiting the Many-Body Localization (MBL) transition. 
 Inspired by the analysis of stability in dynamical systems, this crossover is presented as a flow pattern in the parameter space, with the Poisson statistics being the fixed point of the system, which represents the MBL phase.
We also analyze an isotropic Heisenberg spin-$1/2$ chain with disordered local exchange coupling and a zero magnetic field. In this case, the system never achieves the MBL phase because of the spin rotation symmetry. 
This case is more sensitive to finite-size effects than the previous one, and thus the flow pattern resembles a two-dimensional random walk close to its fixed point. 
We propose a system of linearized stochastic differential equations to estimate this fixed point. 
We study the continuous-state Markov process that governs the probability of finding the system close to this fixed point as the disorder strength increases. 
In addition, we discuss the conditions under which the stationary probability distribution is given by a bivariate normal distribution.
%%We discuss in what conditions the stationary probability distribution to find the system close to this fixed point as the disorder strength increases is given by a bivariate normal distribution.
%% The stationary probability distribution to find the system close to this fixed point as the disorder strength increases is obtained by finding the the fixed point of a continuous-state Markov process.
%%The proposed expression effectively describes the MBL transition and can potentially be applied to the study of the spectral properties of other complex networks.
%\textcolor{magenta}{We show that our fitted parameters can be used to determine the critical field in which the MBL transition takes place.}
%%% Critical disorder analysis %%%
%\textcolor{magenta}{Finally, we use the best-fit parameter as a function of the disorder strength in the disordered local field case obtained from our proposed expression to determine the critical field in which the MBL transition takes place.} 
\end{abstract}

\maketitle

\section{\label{sec:level1} Introduction}

The statistical properties of complex system spectra have attracted significant interest over the last century, largely due to the success of Wigner's random matrices theory (RMT) proposed in the 1950s \cite{wigner_rmt:1955} to study the spectrum of a heavy-atom nucleus. RMT provides a powerful framework for understanding the statistical behavior of eigenvalues in large complex systems \cite{mehta2004random}.
%, with the semicircle law being one of its key results, predicting the distribution of eigenvalues in the bulk of the spectrum for large random matrices.
An illustrative example of RMT's application is in the statistical analysis of the spacing between two consecutive energy levels $s$ of many-body Hamiltonians. 
%\textcolor{blue}{in the statistical analysis of the gap between two consecutive energy levels $s$ of many-body Hamiltonians.}
%where 
The Wigner surmise for the classical ensembles of RMT, namely the Gaussian orthogonal ensemble (GOE), the Gaussian unitary ensemble (GUE) and the Gaussian symplectic ensemble (GSE), provides an accurate approximation for the probability distribution $P(s)$, highlighting the universal nature of spectral fluctuations in complex quantum systems. 

The unfolding eigenvalues procedure is a crucial step in analyzing level spacing statistics. 
This procedure removes the global spectral density,
which is system-dependent, and allows for a detailed study of the local spectral properties of complex systems. 
Since the analytical form of the global spectral density is generally unknown, most of the unfolding procedures rely upon a polynomial fitting to the density of the eigenvalues. This procedure can be tricky, as the statistical quantities depend on the method used to unfold the spectrum \cite{gomez_etal:pre2002,ABULMAGD:PhysA2014,bertrand_garcia:prb2016,ABUELENIN:PhysA2018,torres_vargas_etal:pre2018}.

% Paragraph 2: Ratio of consecutive level spacing 

%\newparag{To circumvent the unfolding eigenvalues procedure needed to study the level spacing statistics, Oganesyan and Huse proposed the ratio of consecutive level spacing distribution.}{EXPLAIN}{EXAMPLE: MBL}{SUMMARIZE}

To avoid the unfolding eigenvalues procedure, 
%needed to study the \textcolor{blue}{level spacing statistics,} 
Oganesyan and Huse \cite{Oganesyan:2007} proposed the 
%ratio of consecutive level spacing distribution. 
consecutive-gap ratio $r$.
The probability distribution $P(r)$ provides a scale-independent metric that effectively captures spectral correlations without requiring the global spectral density, making it a robust alternative to traditional level spacing analysis. Atas et al. \cite{Atas:PRL2013} further developed a surmise expression for $P_{\beta}(r)$ that describes the classical ensembles of RMT, where $\beta$ is equal to 1 for GOE, 2 for GUE and 4 for GSE.
Even though this expression is only exact for the simplest system containing only three eigenvalues, it fits very well the 
%ratio of consecutive level spacing 
consecutive-gap ratio distribution
%probability 
for the three classical RMT ensembles even for larger systems, as shown in Ref. \cite{Atas:PRL2013}.
%,with applications ranging from many-body systems to the zeros of the Riemann zeta function.
Although the 
%Wigner-like surmise expression 
proposed surmise in Ref. \cite{Atas:PRL2013} is very useful for many systems described by pure ensembles: there are numerous instances where we are interested in systems undergoing a crossover between these pure ensembles \cite{kumar_pandey:pre2009, kumar_pandey:jpa2010,allez_etal:prl2012, sarkar_etal:pre2020} as parameters like disorder or energy are varied.
Such systems are described by crossover ensembles that exhibit statistical properties that transition between different ensembles, reflecting the complex behaviors characteristic of many-body systems.
%undergoing this crossover.
%}

% Paragraph 3: GOE-Poisson Transition

%\newparag{STATES}{EXPLAIN, EXPAND, SUPPORT EVIDENCE}{EXAMPLE}{SUMMARIZE, TRANSITION}

The crossover between GOE and Poisson statistics illustrates the crucial role of crossovers in understanding spectral transitions in complex systems. This crossover describes how the statistical properties of eigenvalues evolve as a system transitions from chaotic to integrable regime. In the GOE, eigenvalues exhibit strong repulsion, 
%characterized by Wigner-Dyson statistics, 
while in Poisson statistics, typical of integrable systems, eigenvalues are uncorrelated.
%, leading to a different spacing distribution. 
An example of this crossover can be seen in the many-body localization (MBL) transition \cite{BASKO20061126} in isolated interacting quantum many-body systems, where the system moves from an ergodic phase, described by GOE, to a localized phase, described by Poisson statistics, as the disorder strength increases \cite{Oganesyan:2007}. 
%\textcolor{red}{Understanding this crossover is crucial for characterizing the spectral properties of complex quantum systems}
% CARLOS' COMMENT: Repetion of the first sentence.
%\textcolor{blue}{and sets the stage for our proposed surmise expression for the ratio of consecutive level spacing distribution in such scenarios.}
% CARLOS' COMMENT: save it for the next paragraph, altough some in this line should be hinted here.
% FIRST PARAGRAPH OF SUBSECTION IIC
%The main goal of this paper is to propose a generalized version of this surmise expression to investigate systems described by \textcolor{orange}{crossover ensembles}, which display statistical properties transitioning between pure ensembles. We are mainly interested in the GOE ($\beta=1$) to Poisson distribution 
%transition 
%\textcolor{orange}{crossover}, which is believed to occur in the MBL phase transition as the system \textcolor{orange}{disorder strength} increases \cite{Pal:2010}. 
%SEE THIS AGAIN:\textcolor{red}{Maybe it is a good idea to comment the GOE-GUE crossover and cite the paper Sarkar et al. PRE 101 012216 (2020).}

% Paragraph 4: General distribution. 

%\newparag{In this paper, we propose a Brody-like distribution (BLD), characterized by a parameter, $0<b<1$, for the ratio of two consecutive level spacing of the spectrum of many-body Hamiltonians.}{EXPLAIN}{EXAMPLE}{SUMMARIZE}

%%\textcolor{magenta}{
%The main goal of this paper is to investigate systems described by 
%%\textcolor{orange}{
%crossover ensembles,
%%} 
%which display statistical properties transitioning between pure ensembles.
%%}
This paper proposes a surmise expression for the 
%ratio of consecutive level spacing distribution 
consecutive-gap ratio distribution
$P_{\beta, \gamma}(r)$ characterized by parameters $\beta$ and $\gamma$ for systems that undergo a GOE-to-Poisson statistics crossover.
The parameter $\beta$ is equivalent to the Dyson parameter, ranging from zero to one, and $2-\gamma$ plays the role of an effective confining potential, with $\gamma$ varying from zero to two. 
Surmise expressions such as $P_{\beta, \gamma}(r)$ have already been proposed in the literature \cite{Corps:PRB2020,KARAMPAGIA:NPA2022}.
Nevertheless, our present work %advances in proposing 
is distinctive in that it proposes an expression inspired by the 
%one-dimensional \textcolor{red}{logarithmic gas} picture of the eigenvalue repulsion 
eigenvalues joint probability distribution of the classical ensembles of RMT \cite{mehta2004random,Forrester:book2010}.
%The parameter $\beta$ is equivalent to the Dyson parameter, ranging from zero to one, and $2-\gamma$ plays the role of an effective confining potential, with $\gamma$ varying from zero to two. 

%Here 
We apply the proposed distribution $P_{\beta, \gamma}(r)$ to study 
%the statistical properties of the ratio of consecutive level spacing 
consecutive-gap ratio statistics of 
%an isotropic Heisenberg spin-$1/2$ chain where we consider two distinct cases, namely, the {\it local field case}, with constant exchange coupling and disordered local field, and the {\it exchange coupling case}, with disordered local exchange coupling and zero magnetic field, employing the exact diagonalization method for chain lengths up to 18 spins. 
%We study 
an isotropic Heisenberg spin-$1/2$ chain with two types of disorder: (i) the local field disorder, characterized by constant exchange couplings and a disordered local magnetic field, and (ii) the exchange coupling disorder, involving disordered exchange couplings and a zero magnetic field. 
We use exact diagonalization to compute the eigenvalues of spin chains with lengths up to 18 sites.
%\textcolor{red}{The Hamiltonian of the former case is invariant under global $SU(2)$-rotations.}
Varying the maximal local field strength $h$ in the local field case (i) or the maximal exchange coupling strength $b$ in the exchange coupling case (ii) allows us to detect signs of the MBL transition in the ensemble-averaged 
%ratio of consecutive level spacing distribution 
consecutive-gap ratios,
%obtained from numerical data, 
demonstrating strong agreement between our proposed two-parameter expression and the numerical data. 

% Three-stage scheme

%\textcolor{red}{Inspired by the two-stage idea proposed in Ref. \cite{Serbyn:PRB2016}, we described the GOE-to-Poisson statistics crossover in three stages based on the parameters of $P_{\beta,\gamma}(r)$. 
%We show that increasing $h$ causes the averaged consecutive-gap ratios to transition from GOE to Poisson statistics, passing through all stages. In contrast,  the system remains in the second stage upon increasing $b$.
%We propose that the concept of stages offers a unified framework for understanding and classifying the intermediate behaviors of distinct models during their transition from an ergodic to a localized phase. This framework has the potential to shed light on the spectral properties of quantum many-body systems and the dynamics of real-world complex networks.}

% Flow patterns
By drawing a parallel with dynamic systems, the 
%crossover from GOE to Poisson statistics 
GOE-to-Poisson statistics crossover
is depicted as 
%flow patterns 
traced trajectories
within the parameter space,  which progresses as $h$ or $b$ increases. 
%within both chains. 
For the disordered local field case, the fixed point is the MBL phase characterized by the Poisson statistics. 
The disordered exchange coupling case seems to be more sensitive to the size of the chain and the number of disorder realizations considered in the ensemble averaging since its 
%flow pattern
traced trajectory
resembles a two-dimensional Brownian motion as $b$ increases. To find out the nature of the fixed point and its stability, we propose a system of stochastic differential equations (SDEs) formed by a linearized deterministic part that controls the nature of the fixed point and an additive white noise playing the role of the fluctuation of the parameters due to finite length of the chain and size of the ensemble of disorder realizations effects. 

We find that the fixed point in this case is associated with a non-ergodic phase distinct from the MBL phase,
in agreement with Ref. \cite{Potter:PRB2016}, which states that non-abelian symmetries, such as the global symmetry $SU(2)$, see, e.g.,  Ref. \cite{protopopov_etal:prb2017}, prevent the emergence of a localized phase, even for high values of the maximal exchange coupling strength.
%\textcolor{blue}{We further show that the stationary probability distribution to find the system close to this fixed point as $b$ increases, namely, the fixed point of a continuous-state Markov process, is given by a bivariate normal distribution assuming that the transition probabilities are normally distributed.}
We further show that the stationary probability distribution, which describes the probability of the system being close to this fixed point as the parameter $b$ increases, specifically the fixed point of a continuous-state Markov process, is characterized by a bivariate normal distribution, assuming that the transition probabilities are normally distributed.

\begin{comment}

\textcolor{red}{
We also show that the parameters fitted from our proposed expression can be used to determine the maximal local field critical strength $h_{\rm c}$ where the MBL transition takes place. 
The value estimated for us is in agreement with the ones obtained in Refs. \cite{Schliemann:PRB2021,DeLuca_Scardicchio:EPL2013}, 
%analyzing the averaged consecutive-gap ratio %$\braket{r}$ 
%together with its sample-to-sample variance, 
%$(\Delta_s r)^2$, 
although slightly lower than those commonly cited in the literature for this transition \cite{Pal:2010,Luitz:PRB2015}.
}

\end{comment}

%\newparag{This paper is organized as follows}{}{}{}

This paper is structured as follows. Section \ref{sec_ratiodist} discusses the 
distribution to examine the 
%crossover between the GOE and Poisson statistics. 
GOE-to-Poisson statistics crossover.
Section \ref{sec_spinchains} covers the spin model and details the numerical methods used. 
Numerical results and analyses are presented in Section \ref{sec_numerics}. Finally, our main conclusions are summarized in Section \ref{sec_conclusion}.

%\section{Ratio of consecutive level spacing distribution}
\section{Consecutive-gap ratio distribution}
\label{sec_ratiodist}

The focus of this study is on the statistical measure known as 
%the restricted 
%ratio of consecutive level spacing 
consecutive-gap ratio
defined as \cite{Oganesyan:2007}

\begin{equation}
r_n \equiv
%\frac{\min \left\{s_{n+1}, s_{n}\right\}}{\max \left\{s_{n+1}, s_{n}\right\}}
%=
\min \left\{\tilde{r}_n, \frac{1}{\tilde{r}_n}\right\},
%\, \tilde{r}_n=\frac{s_{n+1}}{s_{n}},
\end{equation}
where $\tilde{r}_n=s_{n+1}/s_{n}$ and $s_n=e_{n+1}-e_n$ is the spacing between two consecutive eigenvalues $e_{n+1}>e_n$ of a given Hamiltonian $H$. 
Our goal is to analyze the probability distribution $P(r)$ for a broad category of many-body systems.

%\subsection{Ratio of consecutive level spacing distribution for Wigner-Dyson Ensembles}
\subsection{Consecutive-gap ratio distribution for Wigner-Dyson Ensembles}
\label{WD_ratiodist}

In Ref. \cite{Atas:PRL2013}, the authors have proposed a surmise expression $P_{\beta}(r)$ for the three Wigner-Dyson ensembles, where $\beta=1,2$ and $4$ represent GOE, GUE and GSE, respectively. 
This expression 
%\textcolor{red}{
%which here we refer to $P_{\beta}(r)$,
%}
is inspired by the Wigner surmise %expression 
for the level spacing statistics \cite{mehta2004random}, which is exact for the $2 \times 2$ random matrices.
%\textcolor{red}{ 
%where $\beta=1,2$ and $4$ represent GOE, GUE and GSE, respectively.
%}
The surmise expression proposed in Ref. \cite{Atas:PRL2013} is obtained from 
%the 
%simplest case, a three-level system characterized by its 
%probability of consecutive level spacing $P(s_1,s_2)$, as follows     

%which can be obtained from the joint probability of two consecutive level-spacing, $P(s_1,s_2)$ as follows:  

%\begin{equation}
%P_{\beta}(r) \equiv 2 \Theta (1 - \tilde{r})\int P_{\beta}\left(s_1, s_2\right) \delta\left(\tilde{r}-\frac{s_1}{s_2}\right) d s_1 d s_2,
%\label{ratio_level_dist}
%\end{equation}

\begin{equation}
P_{\beta}(r) \equiv 2 \int^{\infty}_0\int^{s_2}_0 P_{\beta}\left(s_1, s_2\right) \delta\left(r-\frac{s_1}{s_2}\right) \mathrm{d}s_1 \mathrm{d}s_2,
\label{ratio_level_dist}
\end{equation}
%where the restricted distribution is related to Eq. (\ref{ratio_level_dist}) by the simple relation $P(r) = 2 P(\tilde{r}) \Theta (1 - \tilde{r})$, where $\Theta(x)$ is the Heaviside function.
%where $P(s_1,s_2)$ is the joint distribution of two consecutive level-spacings, which can be obtained from the joint probability density function (jpdf):  
where 
%$\Theta(x)$ is the Heaviside function and 
$P_{\beta}(s_1,s_2)$ 
%is the two consecutive level spacing probability. 
is the joint probability distribution of two consecutive level spacing.
For the simplest case, namely a $3 \times 3$ matrix with eigenvalues $e_1 < e_2 < e_3$ obtained from 
%the eigenvalue probability density function 
the joint probability of $N$ eigenvalues
for the Wigner-Dyson ensembles \cite{mehta2004random}, 
%\textcolor{magenta}{
\begin{equation}
\rho_{\beta}\left(e_1, \ldots, e_N\right)
%=C_{\beta}(N) 
\propto
\prod_{1 \leq i<j \leq N}\left|e_i-e_j\right|^\beta \prod_{i=1}^N e^{-e_i^2 / 2},
\label{WD-LogGas}
\end{equation}
%this probability reads
%where $C_{\beta}(N)$ is a normalization constant 
%for $N$ eigenvalues,
%. For $N=3$ 
$P_{\beta}(s_1,s_2)$ is given by
%and

\begin{equation}
    P_{\beta}(s_1,s_2) \propto s_1^{\beta} s_2^\beta(s_1+s_2)^\beta e^{-\frac 16\left[s_1^2+s_2^2+(s_1+s_2)^2\right]} .
    \label{joint_spacing_RMT}
\end{equation}
%where $C_{\beta}$ is a normalization constant. 
Substituting Eq. (\ref{joint_spacing_RMT}) in Eq. (\ref{ratio_level_dist}) and performing the integrals, one obtains the surmise 
%as mentioned above 
expression
%, 

\begin{equation}
P_{\beta}(r) = Z^{-1}_{\beta} \frac{r^{\beta}(1+r)^{\beta}}{\left(1+r+r^2 \right)^{1+3\beta/2}}, 
\label{WD_ratio_dist}
\end{equation}
with $Z_{\beta}$ being a normalization constant.
This expression is in good agreement with numerical data and the exact result for large GUE 
%($\beta = 2$) 
matrices as shown in Ref. \cite{Atas:PRL2013}.

\subsection{Consecutive-gap ratio distribution for the GOE-to-Poisson statistics crossover}
\label{Cross_ratiodist}

The main goal of this paper is to propose a surmise expression to investigate systems described by crossover ensembles.
We are mainly interested in the GOE-to-Poisson statistics crossover, which is believed to occur in the MBL phase transition as the system's disorder strength increases \cite{Pal:2010}. One-parameter expressions for $P(r)$ were recently proposed in Refs. \cite{Corps:PRB2020, KARAMPAGIA:NPA2022}. 
Reference \cite{Corps:PRB2020} proposes a novel expression, initially formulated with two parameters. Through a skillful application of an {\it ansatz}, the authors simplify this initial expression into a one-parameter form.
On the other hand, Ref. \cite{KARAMPAGIA:NPA2022} introduces a distinct one-parameter expression reminiscent of the Brody distribution for the level spacing distribution \cite{Brody:1973}. 

Despite the commendable performance of the expressions mentioned above in capturing the GOE-to-Poisson statistics crossover, our paper advances a compelling third option that, in our assessment, presents a more fitting and nuanced description of this crossover.
%Considering 
%the surmise expression, Eq. (\ref{WD_ratio_dist}) 
Inspired by Eq. (\ref{joint_spacing_RMT}), obtained from the RMT joint probability distribution Eq. (\ref{WD-LogGas}),
%\textcolor{red}{and the two-parameter distribution Eq. (\ref{spacing_Gen_dist})}, 
and the two-parameter expression for the level spacing distribution presented in Ref. \cite{Serbyn:PRB2016},
we propose the following expression for the probability of two consecutive level spacings:  

\begin{equation}
    P_{\beta,\gamma }(s_1,s_2) \propto  s_1^{\beta} s_2^\beta(s_1+s_2)^\beta 
    %e^{-\left[s_1^{2-\gamma}+s_2^{2-\gamma}+(s_1+s_2)^{2-\gamma}\right]} ,
    e^{-A_{\gamma}(s_1,s_2)} ,
    \label{joint_spacing_Ansatz}
\end{equation}
where $A_{\gamma}(s_1,s_2)\equiv s_1^{2-\gamma}+s_2^{2-\gamma}+(s_1+s_2)^{2-\gamma}$, $0 \leq \beta \leq 1$ and $0 \leq \gamma < 2$.

\begin{comment}

%% remove this sentence in red! %%
\textcolor{red}{Unlike the equivalent expression, Eq. (\ref{joint_spacing_RMT}), derived in Ref. \cite{Atas:PRL2013} for Wigner-Dyson ensembles, which is exact for $3 \times 3$ matrices, Eq. (\ref{joint_spacing_Ansatz}) here
is only an approximation for the $3 \times 3$ matrices case.} 
This 
% expression is motivated by a 
follows from generalization of the eigenvalue probability density function, Eq. (\ref{WD-LogGas}), specifically for crossover ensembles
\begin{equation}
\rho_{\beta, \gamma}\left(e_1, \ldots, e_N\right)
%=C_{\beta,\gamma}(N) 
\propto
\prod_{1 \leq i<j \leq N}\left|e_i-e_j\right|^\beta \prod_{i=1}^N e^{-e_i^{2-\gamma} / 2}.
\label{Gen_LogGas}
\end{equation}
%where, $C_{\beta,\gamma}(N)$ is a normalization constant. 
%\textcolor{red}{This is the regular RMT log-gas}
This is the RMT joint probability of $N$ eigenvalues, Eq. (\ref{WD-LogGas}), with a generic confining potential $e_i^{2-\gamma}$. %\textcolor{red}{Of course,} 
It is not possible to derive an exact form for $P_{\beta,\gamma }(s_1,s_2)$ from Eq. (\ref{Gen_LogGas}), since the exponent of the confining potential $2-\gamma$ is no longer an integer. Nevertheless, considering Eq. (\ref{joint_spacing_Ansatz}) we are able to obtain a surmise expression for the %ratio of consecutive level spacing 
consecutive-gap ratio
distribution that very well describes the GOE-to-Poisson statistics crossover.

\end{comment}

Substituting Eq. (\ref{joint_spacing_Ansatz}) in Eq. (\ref{ratio_level_dist}) and performing the integrals, we obtain 

\begin{equation}
    P_{\beta,\gamma}(r) = Z^{-1}_{\beta, \gamma} \frac{r^{\beta}(1+r)^{\beta}}{\left[1+r^{2-\gamma}+(1+r)^{2-\gamma}\right]^{\frac{3\beta+2}{2-\gamma}}},
    \label{General_ratio_dist}
\end{equation}
%where $0\leq \beta \leq 1$, $0 \leq \gamma < 2$. 
where $Z_{\beta, \gamma}$ is a normalization constant.
%, $0 \leq \beta \leq 1$ and $0 \leq \gamma < 2$. 
It is easy to verify that the above expression obeys the relation,
$P(r) = 1/r^2 P \left( 1/r\right),$
%since the distributions for $r_n$ and $1/r_n$ are the same 
due to the $P(s_1,s_2) = P(s_2,s_1)$ symmetry \cite{Atas:PRL2013}. It is also straightforward to check that for $\gamma = 0$ one recovers Eq. (\ref{WD_ratio_dist})
%, $P_{\beta,\gamma = 0}(r) = P_{\beta}(r)$, 
while for $\beta = 0$ and $\gamma = 1$ one obtains the Poisson statistics, $P_{\beta = 0,\gamma = 1}(r) \propto 1/(1+r)^2$. For $r \rightarrow 0$, $P_{\beta,\gamma}(r \rightarrow 0) \propto r^{\beta}$ 
%is proportional to $r^{\beta}$, 
and 
%for $\tilde{r} \rightarrow \infty$, 
$P_{\beta,\gamma}(\tilde{r} \rightarrow \infty) \propto \tilde{r}^{-(2+\beta)}$. 
%is proportional to $\tilde{r}^{-(2+\beta)}$.

%\textcolor{red}{The parameter $\beta$ in Eq. (\ref{General_ratio_dist}) governs the level repulsion similarly to the plasma model outlined in Eq. (\ref{spacing_Gen_dist}), while the parameter $\gamma$, although inspired by $\gamma_{\rm p}$, serves as a phenomenological parameter that, along with $\beta$, characterizes the GOE-Poisson crossover exhibited by the system under investigation.}
%\textcolor{orange}{}
The key finding presented in our work is Eq. (\ref{General_ratio_dist}).
Throughout the remainder of the paper, we demonstrate its effectiveness in characterizing the GOE-to-Poisson statistics crossover by employing it in the analysis of the MBL transition within interacting spin chains.
%\textcolor{red}{Insert here a sentence commenting on our three-stage scheme that will be discussed in the result section.}

\section{Heisenberg spin chains} \label{sec_spinchains}

One-dimensional spin chains play a crucial role in the study of equilibrium and out-of-equilibrium properties of realistic many-body systems. These models serve as an essential testing ground for theoretical concepts because of their relative simplicity and the availability of exact solutions or highly accurate numerical methods.
We consider a spin-$1/2$ Heisenberg chain of length $L$ in the presence of a random Zeeman field as described by the Hamiltonian

\begin{equation}
    H = \sum^{L^{\prime}}_{i=1} J_i \vec{S}_i \cdot \vec{S}_{i+1} + 
    \sum^L_{i=1} h_{z, i} S^z_i,
    \label{General_Heisenberg}
\end{equation}
where $J_i$'s are the couplings between adjacent spins of the chain, and $h_{z, i}$ is the local magnetic field in the $i$-th site of the chain that interacts with the $z$-component of the spin $\vec{S}_i$ and $L^{\prime} = L-1$ ($L$) for open (periodic) 
%chains.
%with periodic, $S_{L+1} = S_1$, 
boundary condition. 

The model presented in Eq. (\ref{General_Heisenberg}) is highly versatile as it does not impose any prior assumptions on the external field $h_{z,i}$ or the exchange couplings $J_i$.
The Heisenberg model has been widely used in numerical studies of MBL transition \cite{Oganesyan:2007,Pal:2010,Huse_etal:2013,Luitz:PRB2015,luitz_etal:prb2016,luitz:prb2016,Schliemann:PRB2021}.
It is important to note that the one-dimensional isotropic Heisenberg model with constant exchange couplings and no random field possesses an exact solution, implying its non-ergodic nature \cite{Baxter_exact:1985,BABUJIAN1982479}.
In the remainder of this section, we discuss some special cases of the general Hamiltonian Eq. (\ref{General_Heisenberg}) with disordered local field and exchange couplings.

\subsection{Isotropic 1D Heisenberg spin chain with constant exchange couplings and disordered local field}\label{subsec_LF}

%\textcolor{red}{
%Perhaps we should first mention that the 1D Heisenberg model has an exact solution with no random field? (non-ergodic).
%}

%\textcolor{blue}{What about something like: ``Before delving into the disordered case, it is important to note that the one-dimensional isotropic Heisenberg model with constant exchange couplings and no random field possesses an exact solution, implying its non-ergodic nature \cite{Baxter_exact:1985,BABUJIAN1982479}.''}

One specific instance of the general Hamiltonian Eq.~(\ref{General_Heisenberg}) involves 
%an open 
a spin-$1/2$ chain experiencing a disordered local field interacting with the $z$-component of each spin. Setting $J_i=J$ and $L^{\prime} = L-1$, Eq.~(\ref{General_Heisenberg}) becomes:
\begin{equation}
    H_{\rm LF} = \sum^{L-1}_{i=1} J \vec{S}_i \cdot \vec{S}_{i+1} + 
    \sum^L_{i=1} h_{z, i} S^z_i,
    \label{Heisenberg_DLF}
\end{equation}
where $h_{z, i} = h_i J$, and $h_i$ represents the local magnetic field interacting with the $i$th spin of the chain, distributed uniformly within the range $[-h, h]$, with $h$ denoting the maximal local field strength.
Our choice to use open boundary conditions in this particular example is arbitrary, as both open and periodic boundary conditions have been widely employed in MBL studies, as seen in works like \cite{Oganesyan:2007,znidaric_etal:prb2008,Huse_etal:2013,Iyer:2013,Luitz:PRB2015,chandran_etal:prb2015,luitz_etal:prb2016,Corps:PRB2020}.

By examining a chain with up to $L=18$ spins and varying the maximal local field strength between $0.1$ and $5.0$, we obtain the 
%ensemble averaged ratio of consecutive level spacing distribution 
averaged consecutive-gap ratios
of the zero magnetization sector (or the minimum magnetization sector for chains comprising an odd number of spins) 
%of the Heisenberg model, as defined in Eq. (\ref{Heisenberg_DLF}).
of this Hamiltonian.
The zero magnetization sector represents a high-energy ``infinite-temperature condition", making it ideal for studying MBL as a global phenomenon that affects the entire energy spectrum, highlighting the transition from thermalization to localization across the full spectrum of many-body systems \cite{Oganesyan:2007}.

This study is carried out using the exact diagonalization, implemented with the package {\tt QuSpin} \cite{QuSpin:2017,QuSpin:2019} for {\tt Python}. 
For each realization in our ensemble, we compute the 
%ratio of consecutive level spacing 
consecutive-gap ratios
and then obtain the averaged distribution of these ratios by performing ensemble averaging. 
We proceed to adjust the parameters of the surmise distribution, Eq. (\ref{General_ratio_dist}), to best fit the ensemble-averaged distribution
by solving a nonlinear least squares optimization problem using the Levenberg-Marquardt algorithm, implemented via the {\tt LMFIT} package for {\tt Python}.

\subsection{Isotropic 1D Heisenberg spin chain with disordered local exchange coupling and zero magnetic field}\label{subsec_LE}

We also investigate a periodic isotropic spin chain characterized by disordered local exchange couplings by setting $h_{z, i} = 0$ and $L^{\prime} = L$ in Eq. (\ref{General_Heisenberg}) 
\begin{equation} 
H_{\rm EC} = \sum^L_{i=1} J_i \vec{S}_i \cdot \vec{S}_{i+1}, 
\label{Heisenberg_DLE}
\end{equation} 
with $J_i = J + b_i$,  $J \geq 0$ and $b_i \in [-b, b]$ distributed uniformly with maximal exchange coupling strength $b$.
As in the local field case, selecting this boundary condition is entirely arbitrary.
This case exhibits 
%\textcolor{red}{a non-abelian symmetry}, as it remains invariant under global $SU(2)$ rotations. 
spin rotation symmetry.
This symmetry is predicted to inhibit the formation of an MBL phase, regardless of how strong the maximal exchange coupling strength $b$ becomes \cite{Potter:PRB2016}. 
Recent numerical investigations have examined disordered spin systems with symmetry $SU(2)$ and identified a type of incomplete MBL phase \cite{protopopov_etal:prb2017,protopopov:prx2020,Siegl:NJP2023}. 
The impact of 
%\textcolor{red}{non-abelian symmetries} 
such symmetry
on the eigenstate thermalization hypothesis (ETH) and entanglement entropies has been explored in recent studies \cite{murthy_etal:prl2023,majidy_etal:prb2023}.

Similarly to the local field case, here we examined chains comprising up to $L=18$ spins.
To address the high degeneracy present in this case, we employ the methodology described in Ref. \cite{Siegl:NJP2023}. 
This approach considers all magnetization sectors $S^z_{\rm tot}$, taking into account all multiplets, with the exception of those with a total spin number $S_{\rm tot} \in \{LS, LS-1\}$. 
After all degeneracies caused by multiplets are removed, the 
consecutive-gap ratios are
calculated from eigenvalues that share the same $S_{\rm tot}$ and $S^{z}_{\rm tot}$.  Once these ratios are computed for a particular disorder realization of the exchange couplings, they are arranged in increasing order before conducting an ensemble average.

\section{Numerical Results} \label{sec_numerics}

Following the methodology presented in section \ref{sec_spinchains}, we compare the %averaged ensemble of the ratio of consecutive level spacing distribution
averaged consecutive-gap ratios
of $L=18$ spin chains described by the Hamiltonians Eqs. (\ref{Heisenberg_DLF})  and (\ref{Heisenberg_DLE}) with the surmise distribution presented in Eq. (\ref{General_ratio_dist}). Our main results are presented in the following subsections.

%\subsection{\textcolor{green}{Ensemble averaged ratio of consecutive level spacing distribution}}
\subsection{Averaged consecutive-gap ratios}
\label{subsec:ratio_dist}

\begin{figure*}[ht]
\centering
\begin{subfigure}{0.225\textwidth}
    \includegraphics[width=\textwidth]{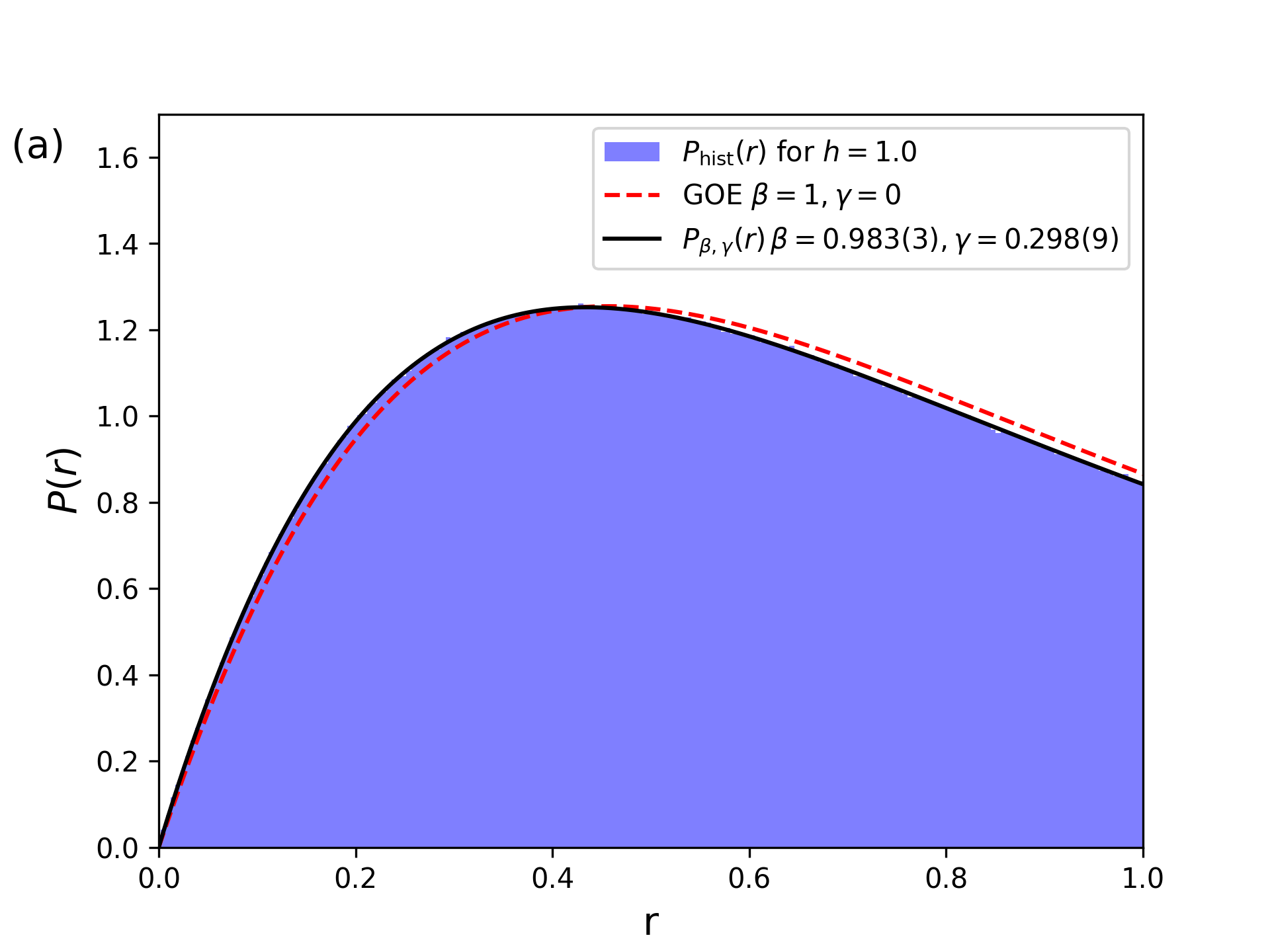}
\end{subfigure}
\begin{subfigure}{0.225\textwidth}
    \includegraphics[width=\textwidth]{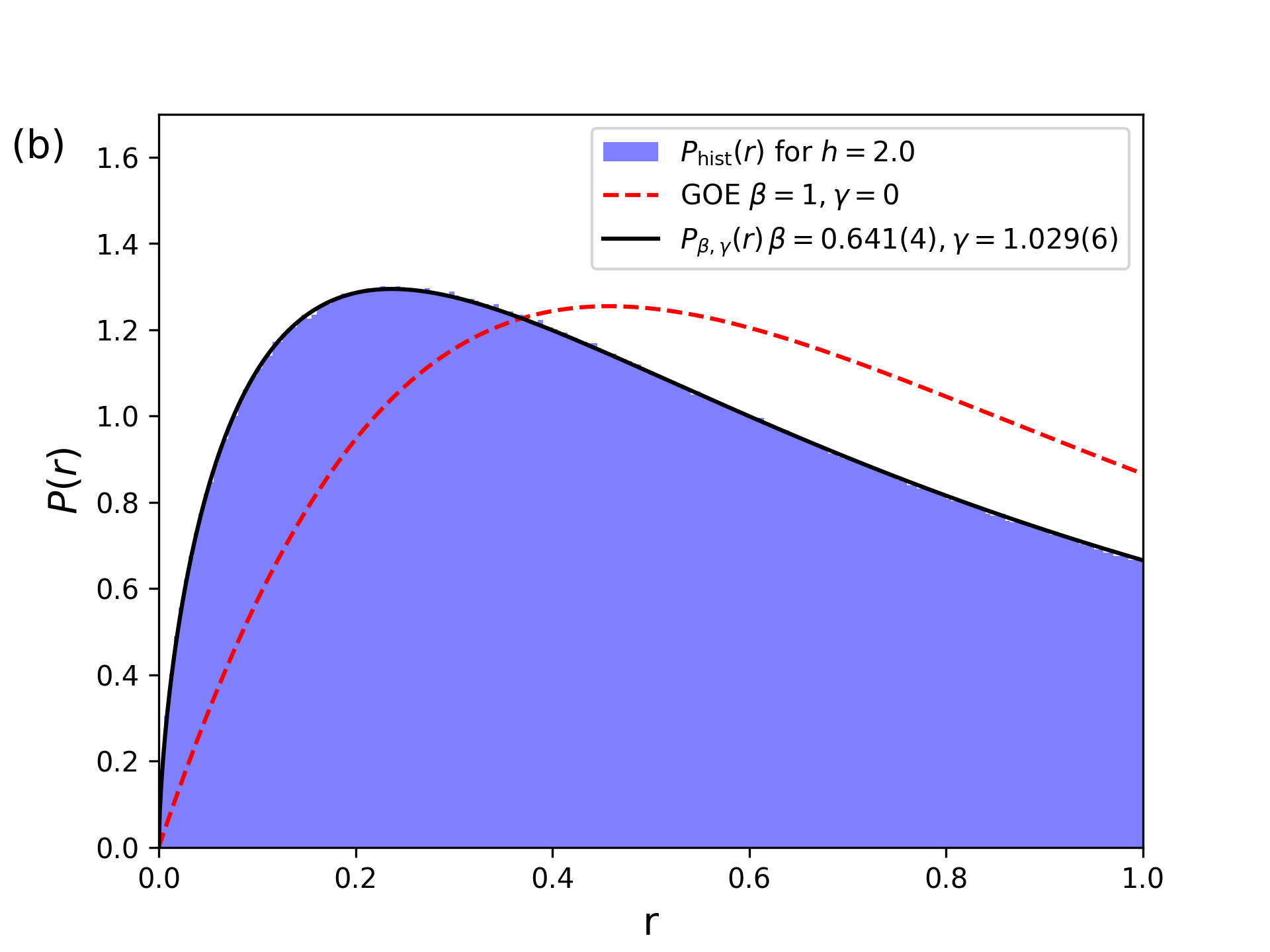}
\end{subfigure}
\begin{subfigure}{0.225\textwidth}
    \includegraphics[width=\textwidth]{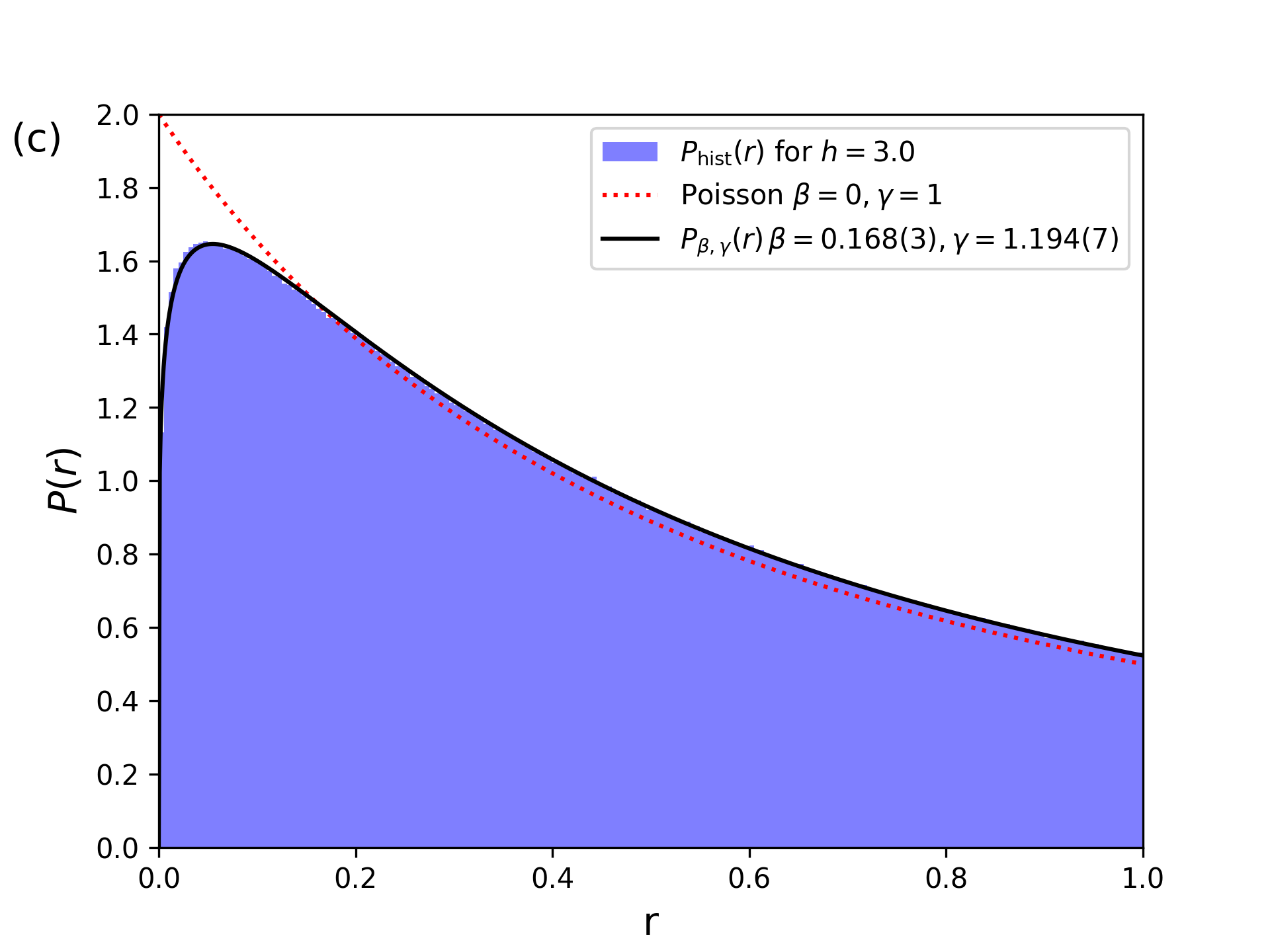}
\end{subfigure}
\begin{subfigure}{0.225\textwidth}
    \includegraphics[width=\textwidth]{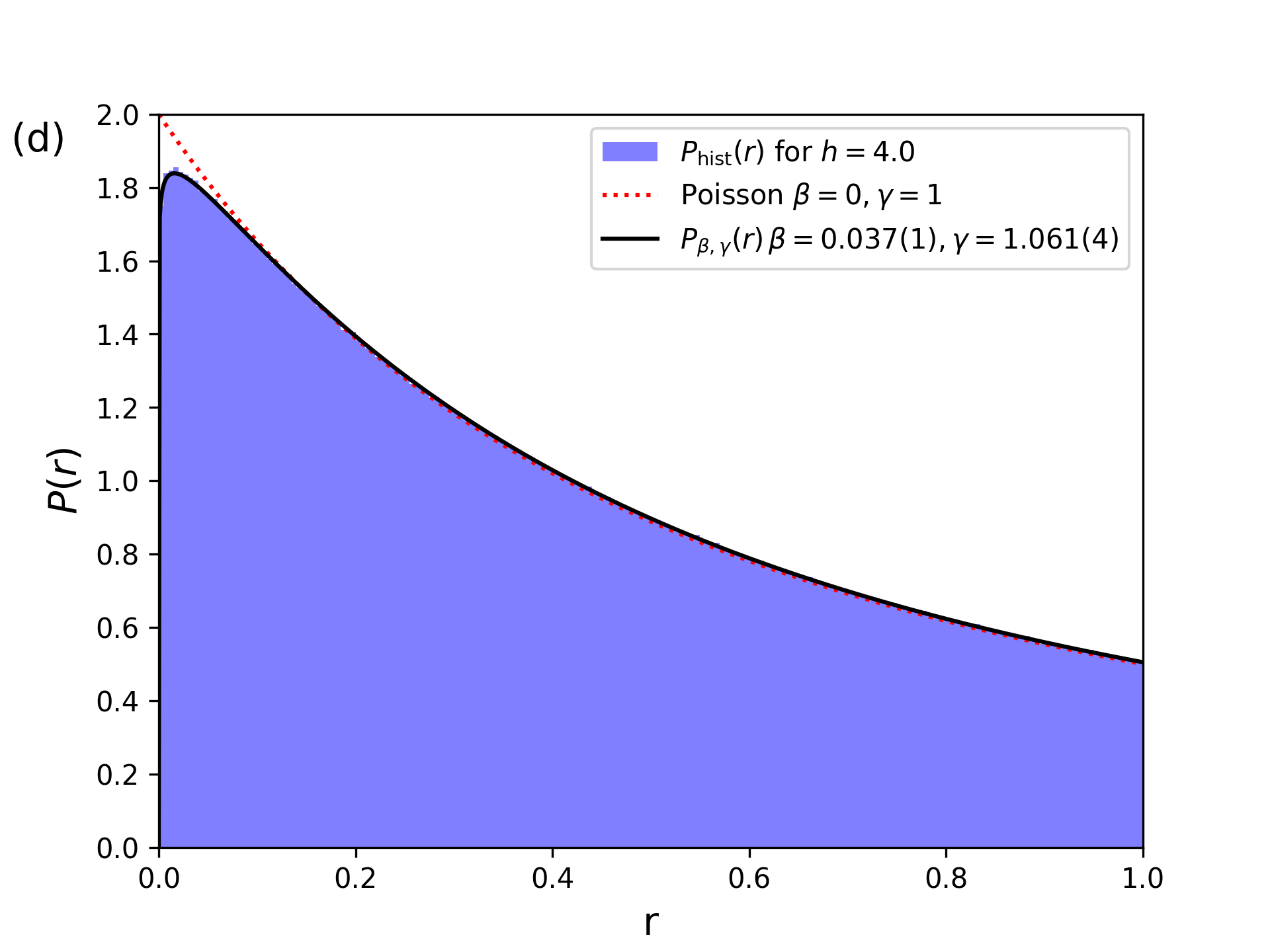}
\end{subfigure}
\begin{subfigure}{0.225\textwidth}
    \includegraphics[width=\textwidth]{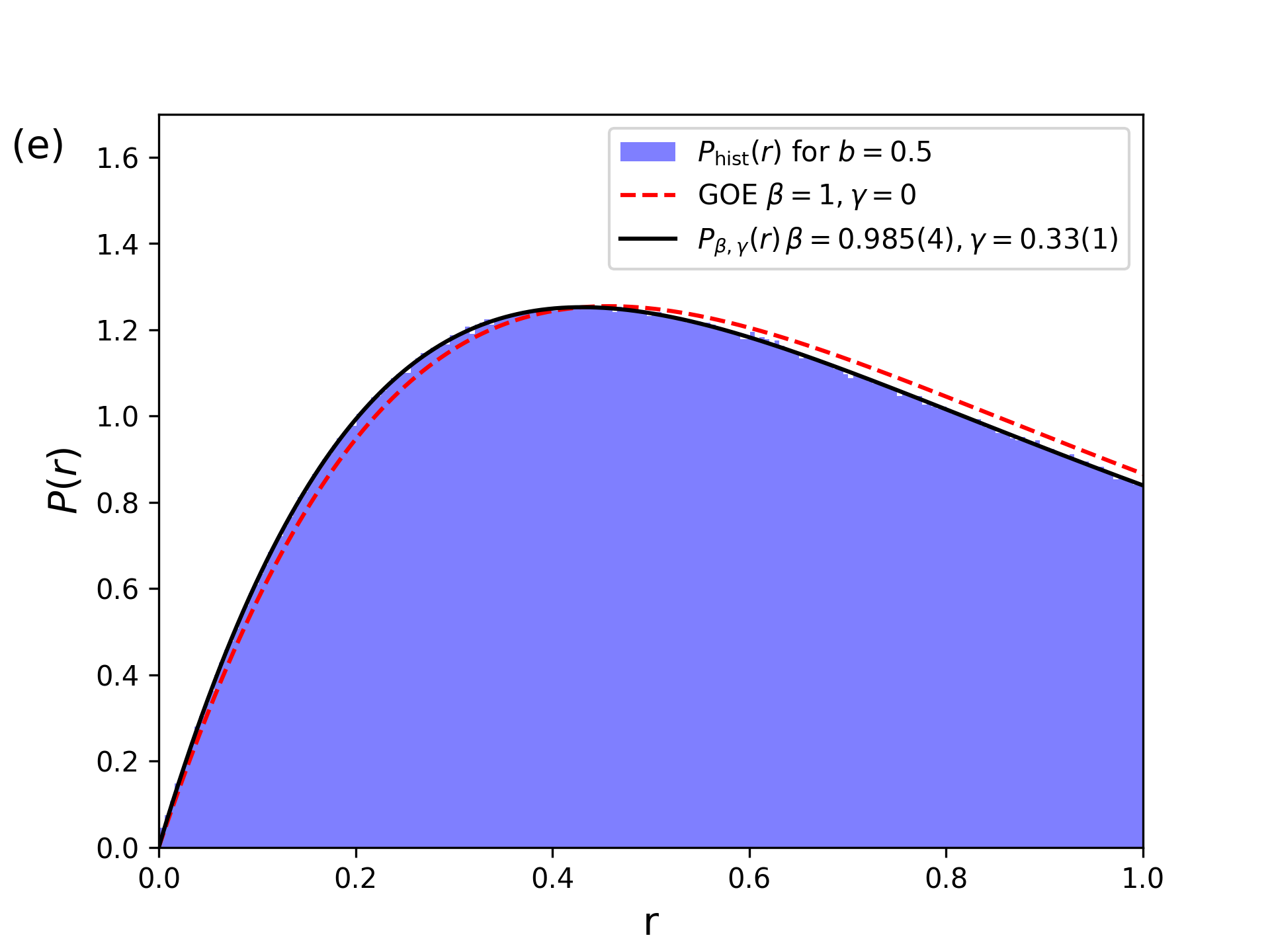}
\end{subfigure}
\begin{subfigure}{0.225\textwidth}
    \includegraphics[width=\textwidth]{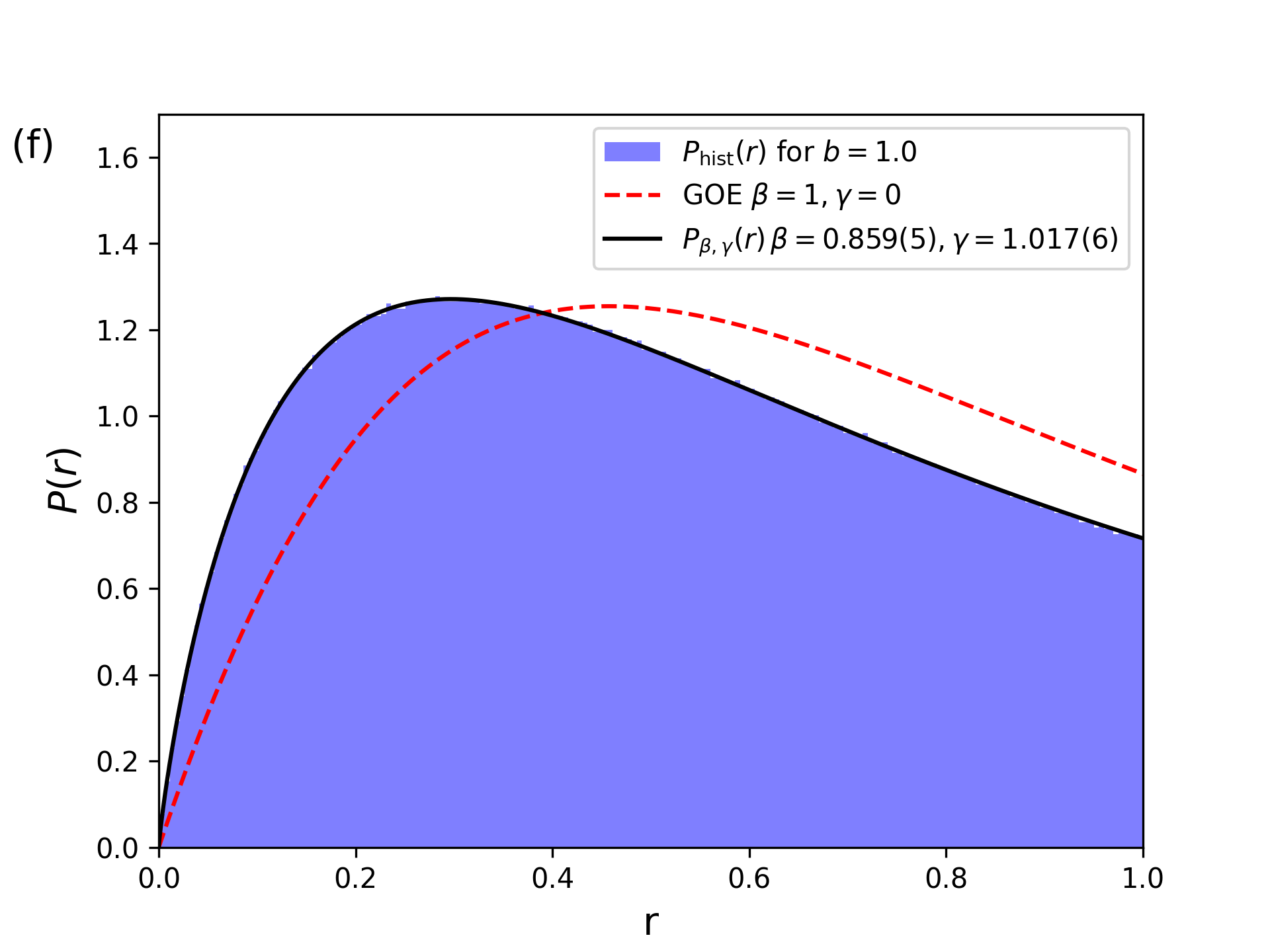}
\end{subfigure}
\begin{subfigure}{0.225\textwidth}
    \includegraphics[width=\textwidth]{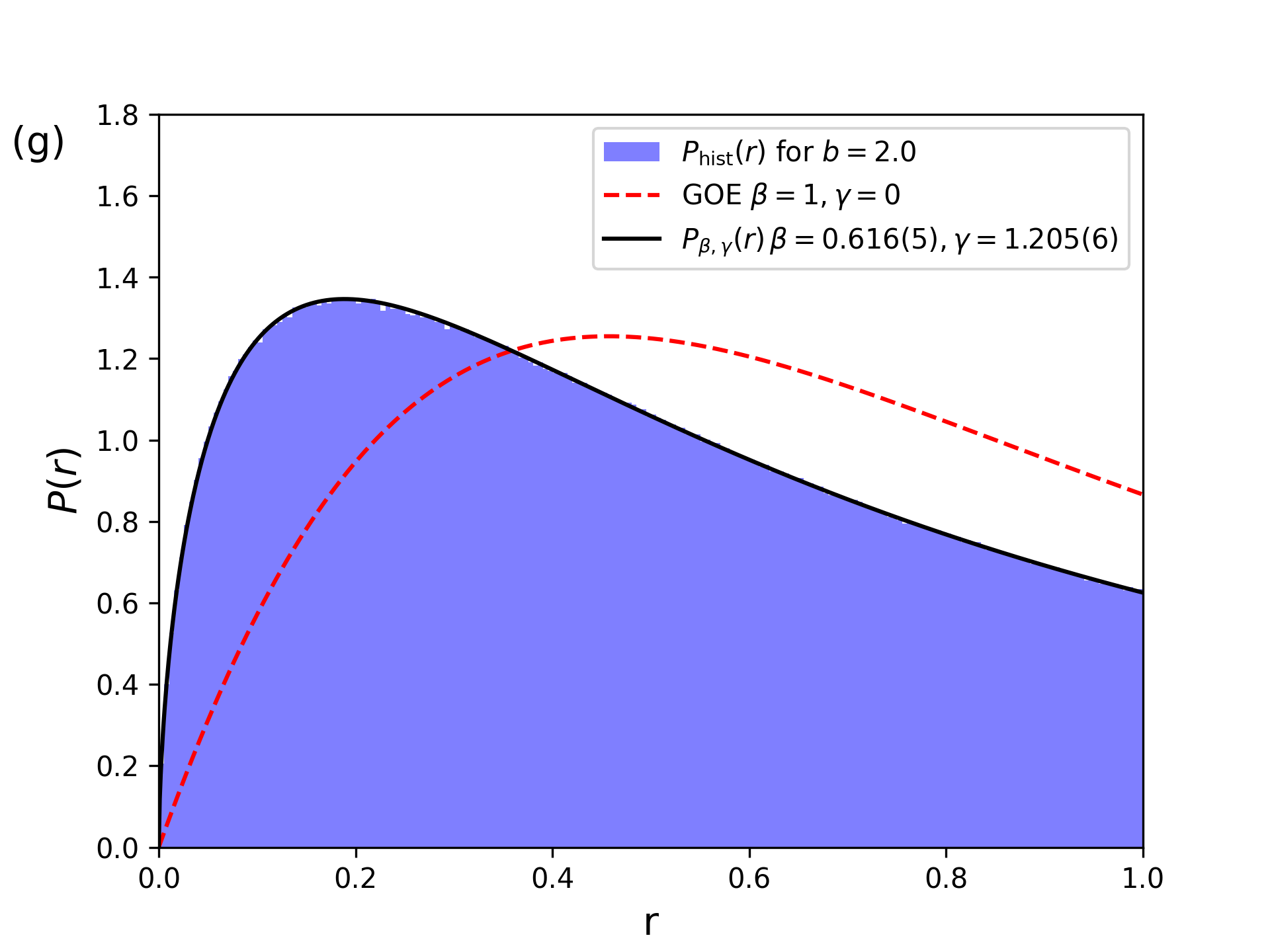}
\end{subfigure}
\begin{subfigure}{0.225\textwidth}
    \includegraphics[width=\textwidth]{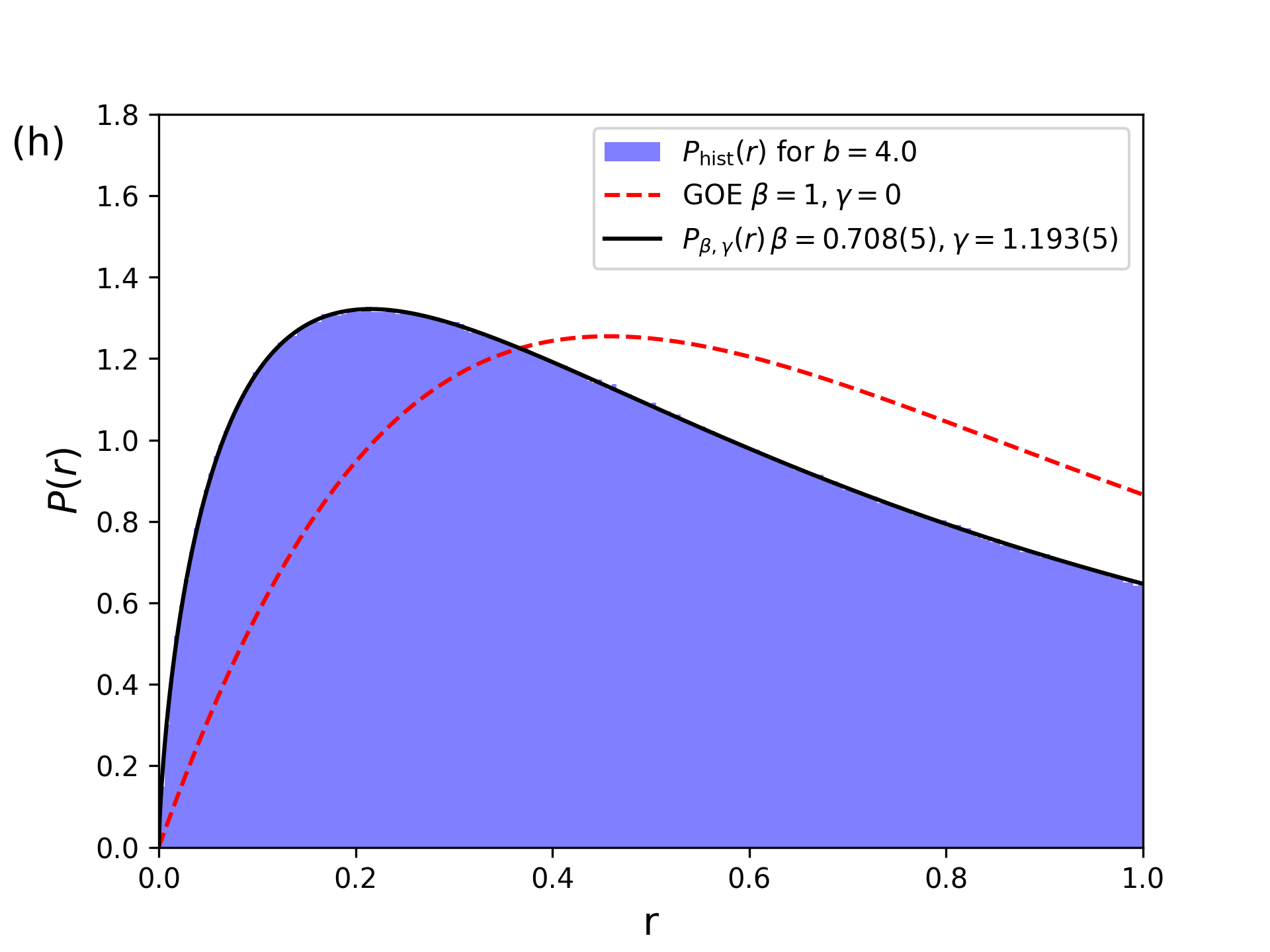}
\end{subfigure}
\caption{
%\textcolor{magenta}{
%Ensemble-averaged ratio of consecutive level spacing distribution 
Averaged consecutive-gap ratios
%$P_{\rm hist} (r)$ 
(blue histograms) for an $L=18$ Heisenberg spin-$1/2$ chain with disordered local field for various 
maximal local field strength
$h$  (upper panels) and disordered exchange couplings 
%for various 
%exchange coupling strengths 
$b$ (lower panels). 
The surmise distribution,
%$P_{\beta,\gamma} (r)$, 
Eq. (\ref{General_ratio_dist}), with best-fit parameters $\beta$ and $\gamma$ for each $h$ and $b$, is represented by a solid black line. 
%The GOE (Poisson) distribution appears in (a) to (d) as dashed red lines and Poisson statistics in (c) to (f) as dotted red lines. 
The GOE [panels (a), (b) and (e)-(h)] and Poisson statistics [(c) and (d)] appear as dashed red lines.
%As depicted in (a), for small values of maximal local field strength, such as $h \lesssim 1.0$, the system is approximately represented by the GOE. 
%Panel (b) depicts the \textcolor{cyan}{first stage} of the crossover with $\beta \approx 0.8$ and $\gamma \approx 0.6$.
%Panel (c)[(d)] represents \textcolor{cyan}{second stage} with $\beta \approx 0.5  (0.6)$ and $\gamma \approx 1.1  (1.2)$. 
%Panel (e) displays \textcolor{cyan}{third stage} with $\beta \approx 0.2$ and $\gamma \approx 1.2$.
%At high levels of local field disorder, $h \gtrsim 4.0$, the statistics align more closely with the Poisson statistics, as shown in (f). 
Upper panels: as $h$ increases, the distribution gradually changes from GOE [panel (a)] to Poisson statistics [panel (d)], indicating that the system reaches the MBL phase.
Lower panels: for small values of $b$, the system is approximately represented by the GOE [panel (e)]. 
%As $b$ increases, the statistical behavior is accurately represented by our surmise expression. 
%\textcolor{red}{Due to the $SU(2)$ symmetry,} 
Due to the spin rotation symmetry the system never reaches the MBL phase. 
This can be seen from panels (f)-(h) as $b$ increases. 
%These distributions resembles the one presented in panel (b), which represents an intermediate stage of GOE to Poisson statistics crossover. 
%}
}
\label{fig:ratio_dists_LF_LE}
\end{figure*}

%Figure \ref{fig:ratio_dists_RLF} illustrates the ensemble-averaged ratio of the consecutive level spacing distribution $P_{\rm hist} (r)$ (depicted by blue histograms), derived from the exact diagonalization of the local field model. The findings are shown for \textcolor{red}{$h = 1.0, 1.7, 2.1, 2.3, 3.0,$} and $4.0$, using a chain consisting of $L=18$ spins, with approximately $4 \times 10^2$ disorder realizations for each given maximal local field strength.

The upper panels of Figure \ref{fig:ratio_dists_LF_LE} illustrate the averaged 
%atio of the consecutive level spacing distribution 
consecutive-gap ratios 
$P_{\rm hist} (r)$, depicted by blue histograms, derived from the exact diagonalization of the local field case 
%(upper panels) 
for $h = 1.0, 2.0, 3.0$ and $4.0$ with approximately $4 \times 10^2$ disorder realizations for each given maximal local field strength $h$.
%, and exchange coupling model (lower panels) for $b = 0.5, 1.0, 2.0$ and $4.0$ with approximately $3 \times 10^2$ disorder realizations per maximal exchange coupling strength $b$.
The solid black curves represent the surmise distribution Eq. (\ref{General_ratio_dist}) using the best-fit parameters derived from the least-squares fitting process. It is evident that the general distribution aligns well with the numerical data.

For low maximal local field strengths, e.g., $h= 1.0$, the distribution closely matches the GOE (red dashed line), as shown in the upper leftmost panel of Fig. \ref{fig:ratio_dists_LF_LE}, even though the estimated value for the parameter $\gamma$ ($\approx 0.3$) is not close to zero. 
For the surmise expression, Eq. (\ref{WD_ratio_dist}) obtained for the classical ensembles \cite{Atas:PRL2013},
%, which is exact for $3 \times 3$, 
we would expect this value to be close to zero. 
As pointed out in Ref. \cite{Corps:PRB2020}, 
%which also proposes a surmise expression for the GOE-Poisson statistics crossover,
the values of these parameters depend on the size of the matrix. 
Therefore, it is not surprising that they will not be exactly the same as those obtained from the prediction of 
%the matrix $3 \times 3$, 
Eq. (\ref{WD_ratio_dist}).

%As the maximal local field strength increases, $h \gtrsim 1.0$, the GOE no longer fits accurately $P_{\rm hist} (r)$. 
%For $h=1.7$, upper central panel of Fig. \ref{fig:ratio_dists_RLF}, with the surmise distribution parameters $\beta = 0.787(3)$ and $\gamma = 0.611(9)$.
%\textcolor{red}{ We employ a classification scheme similar to the one used in Ref. \cite{Serbyn:PRB2016} to describe the crossover in terms of the parameters of Eq. (\ref{General_ratio_dist}). In our classification, the upper central panel of Fig. \ref{fig:ratio_dists_RLF} represents the \textcolor{cyan}{first stage} of the crossover, characterized by $1 \geq \beta \geq 0.5$ and $\gamma < 1$. In this stage, the distribution is qualitatively similar to that of GOE, with the maximum of $\bar{P}_{\beta, \gamma}(r)$ of the same magnitude as that of GOE, but shifted to the left for a lower value of $r$.}

As the maximal local field strength increases, the GOE no longer fits accurately $P_{\rm hist} (r)$, as can be seen in panel (b) of Fig. \ref{fig:ratio_dists_LF_LE} for $h=2.0$. 
%with the surmise distribution parameters $\beta = 0.641(4)$ and $\gamma = 1.029(6)$. 
Panel (c) 
%of Fig. \ref{fig:ratio_dists_LF_LE} 
shows the averaged distribution for $h = 3.0$ 
%using best-fit parameters $\beta=0.168(3)$ and $\gamma = 1.194(7)$. 
%It is 
where is possible to observe that the tail of $P_{\rm hist} (r)$ is already well fitted by the Poisson statistics (dashed red line), indicating that the system is moving towards the localized phase.
With increasing maximal local field strength, the average distribution gradually approaches Poisson statistics, as is evident in the upper right panel of Fig. \ref{fig:ratio_dists_LF_LE} for $h = 4.0$, showing that the system has transitioned to the MBL phase, characterized by $\beta \to 0$ and $\gamma \to 1$.

The lower panels of Fig. \ref{fig:ratio_dists_LF_LE} present 
%the 
%ensemble-averaged distribution of the ratio of consecutive level spacing 
%averaged consecutive-gap ratios
$P_{\rm hist} (r)$ 
%obtained from the exact diagonalization method 
for the exchange coupling case with $b = 0.5,\,1.0,\,2.0$ and $4.0$ for a chain with $L=18$ spins and approximately $3 \times 10^2$ disorder realizations for each maximal exchange coupling strength $b$.
%The solid black lines represent the surmise distribution described in Eq. (\ref{General_ratio_dist}), with parameters optimized through the least-squares fitting method. 
As in the local field case, it is evident that Eq. (\ref{General_ratio_dist}) matches the numerical data precisely. 
As observed in the lower left panel of Fig. \ref{fig:ratio_dists_LF_LE} for $b = 0.5$, the distribution closely resembles the GOE (dashed red line), 
%with $\beta = 0.985(4)$ and $\gamma=0.33(1)$
indicating that the system is in the ergodic phase.

In contrast, the GOE no longer matches the histograms shown in the panels (f), (g), and (h) of Fig. \ref{fig:ratio_dists_LF_LE} for $b=1.0$, $2.0$, and $4.0$, respectively. 
These distributions resemble the one presented in panel (b) for $h=2.0$, which represents an intermediate stage of the GOE-to-Poisson statistics crossover.
Unlike the local field case, the system does not attain the MBL phase, characterized by $\beta \rightarrow 0$ and $\gamma \rightarrow 1$. 
This characteristic of the exchange coupling case aligns with the fact that %\textcolor{red}{the symmetry $SU(2)$}
the spin rotation symmetry
hinders the system from fully entering the localized phase \cite{protopopov_etal:prb2017}.

\begin{figure}[htbp]
    \centering
    \includegraphics[width=0.4\textwidth]{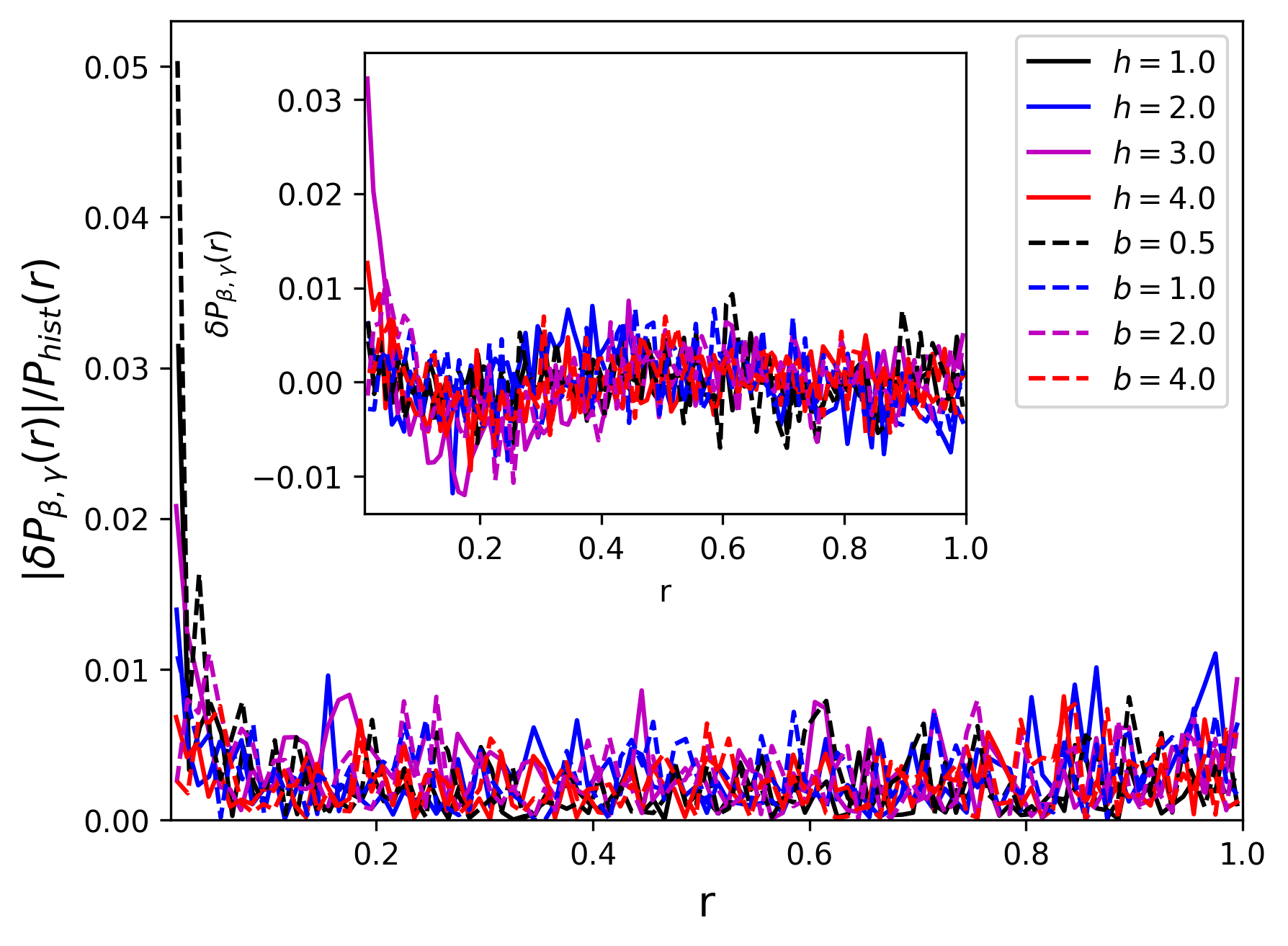}
\caption{Relative error $|\delta P_{\beta,\gamma} (r)| / P_{\rm hist} (r)$ {\it vs.} $r$ for $h=1.0$, $2.0$, $3.0$ and $4.0$ (solid curves) and $b=0.5$, $1.0$, $2.0$ and $4.0$ (dashed curves). The relative error between the surmise distribution, Eq. (\ref{General_ratio_dist}), fitted with the best-fit parameters, and the histogram is minimal. It is approximately $1\%$ for different disorders, except for $r\rightarrow 0$. Inset: error $\delta P_{\beta,\gamma} (r)$ {\it vs.} $r$ for the same values of $h$ and $b$ from the main plot.}
\label{Error_LF_LE}
\end{figure}

Figure \ref{Error_LF_LE} illustrates the error $\delta P_{\beta,\gamma} (r) \equiv P_{\beta,\gamma} (r) -  P_{\rm hist} (r)$ and the relative error $|\delta P_{\beta,\gamma} (r)| / P_{\rm hist} (r) $ for $h=1.0$, $2.0$, $3.0$ and $4.0$ (solid curves) and $b=0.5$, $1.0$, $2.0$ and $4.0$ (dashed curves). 
Observe that the relative error between the surmise distribution Eq. (\ref{General_ratio_dist}) with best-fit parameters and the histogram is of the order of ~$1\%$, 
%for the values considered of $h$ and $b$, 
except for $r\rightarrow 0$, where this relative error approaches ~$3\%$ for $h=1.0$ and ~$5\%$ for $b=0.5$, for example.
This relative error is comparable to the relative error noted in Ref. \cite{Atas:PRL2013} for the surmise expression, Eq. (\ref{WD_ratio_dist}). 
The surmise distribution
%, when fitted using the best-fit parameters, 
is in excellent agreement with the numerical results, and the relative errors are minimal and comparable to those reported in previous studies for the classical ensembles of RMT.
%\textcolor{red}{We also proceed a similar analysis (not shown here) for the local exchange coupling model and find similar behavior as depicted in Fig. \ref{Error_LF_LE} as we vary the maximal exchange coupling strength $b$.}

\subsection{Flow Patterns in the $\beta\gamma$ space}
\label{subsec:flow_patterns}

\begin{figure}[htbp]
    \centering
    \includegraphics[width=0.5\textwidth]{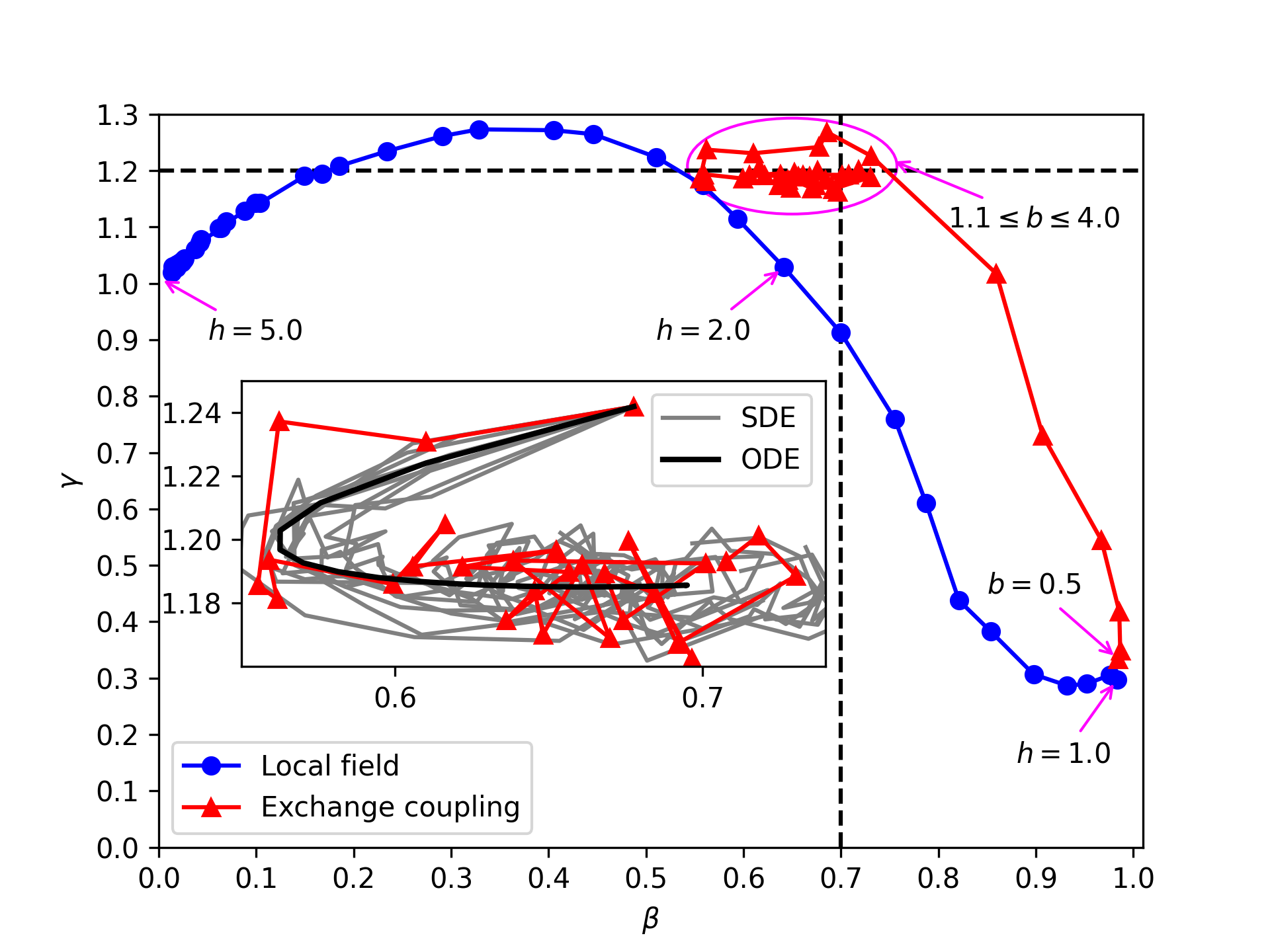}
    \caption{Flow patterns of the local field case (blue) and the exchange coupling case (red) in the $\beta\gamma$ space. 
    %\textcolor{red}{As $h$ increases, the local field case flows toward a fixed point at $\vec{x}^{*}_{\rm LF} = (0,1)^\top$ that characterizes the MBL phase. While when $b$ increases the exchange coupling case seems to be attracted for another fixed point at $\vec{x}^{*}_{\rm EC} \approx (0.7,1.2)^{\rm T}$, represented by the intersection of the horizontal and vertical dashed lines.} 
    As the disorder strength $h$ is systematically increased, the best-fit parameters for the local field case trace a trajectory in the $\beta\gamma$ space, converging towards a fixed point at $\vec{x}^{*}_{\rm LF} = (0,1)^\top$ that characterizes the MBL phase. Similarly, as the exchange coupling disorder $b$ increases, the corresponding best-fit parameters appear to be attracted to another fixed point at $\vec{x}^{*}_{\rm EC} \approx (0.7,1.2)^\top$, represented by the intersection of the horizontal and vertical dashed lines.
    Inset: 
    %zoom at 
    blow up of the $1.1 \leq b \leq 4.0$ region encircled by a magenta ellipse in the main plot. The solid black line represents the best-fit of the data 
    %by the system of SDEs described 
    by Eq. (\ref{eq:Lin-SDE}) with $R=0$. 
    The gray curves represent ten realizations of 
    %the system of SDEs 
    %described by 
    Eq. (\ref{eq:Lin-SDE}) 
    with the best-fit parameters 
    %of the system of ODEs 
    and noise intensity $R=0.02$.}
    \label{fig:Trajectories}
\end{figure}

Fig. \ref{fig:Trajectories} shows the best-fit parameters for the local field case (blue circles) 
%with approximately $4 \times 10^2$ disorder realizations 
for 
%each maximal local field strength 
$h=1.0,\, \ldots,\, 5.0$ in the $\beta\gamma$ space. 
The best-fit parameters for the exchange coupling case (red triangles) are also shown 
%in Fig. \ref{fig:Trajectories} 
%with approximately $3 \times 10^2$ disorder realizations
for 
%each maximal exchange coupling strength 
$b= 0.5,\, \ldots,\, 4.0$. 
%The results presented in Fig. \ref{fig:Trajectories} 
These results can be interpreted as flow patterns in the two-dimensional $\beta\gamma$ space similar to those studied in nonlinear dynamics Ref. \cite{strogatz:book} with the phases seen as fixed points in the $\beta\gamma$ space. Here, the parameters $\beta$ and $\gamma$ play the role of space coordinates, 
%\textcolor{red}{while $h$ and $b$ represent time $t$.} 
while the disorder strengths $h$ and $b$ serve as an analogous progression parameter, akin to time in a dynamical system.
The dynamics is governed by the following system of deterministic differential equations
\begin{equation}
    \dfrac{d \vec{x}}{dt} = \vec{F}(\vec{x}),
    \label{vector_field}
\end{equation}
where $\vec{x} = (\beta(t),\gamma(t))^{\top}$, $\vec{F}(\vec{x}) = (f_{\beta}(\vec{x}),f_{\gamma}(\vec{x}))^{\top}$, $f_{\beta}(\vec{x})$ and $f_{\gamma}(\vec{x})$ are model-dependent functions and 
%$(.)^\top$ stands for 
the superscript $\top$ represents the transpose operation.
%of a vector. 
The fixed points $\vec{x}^{*} \equiv \lim_{t \to \infty}(\beta(t),\gamma(t))^{\top} = (\beta^*,\gamma^*)^{\top}$ are obtained by imposing $\vec{F}(\vec{x}^{*}) = \vec{0}$.

For the local field case, the fixed point is $\vec{x}^{*}_{\rm LF} = (0,1)^{\top}$, which represents the MBL phase described by the Poisson statistics. 
%Performing a linearization procedure in this fixed point... stable node... 
For the exchange coupling case, the fixed point appears to be placed at $\vec{x}^{*}_{\rm EC} \approx (0.7,1.2)^\top$, represented by the intersection of the horizontal and vertical dashed lines in Fig. \ref{fig:Trajectories}.
%.. degenerate star. Brownian motion for $b/J \gtrsim 1.0$
The nature of these fixed points seems to be very different. As $\vec{x}^{*}_{\rm LF}$ looks like a stable node, with both eigenvalues of the linearization matrix around the fixed point being negative, $\vec{x}^{*}_{\rm EC}$ could be a degenerate node, degenerate negative eigenvalues with only one independent eigenvector, or even a stable spiral, when the flow pattern quickly spirals towards the fixed point $\vec{x}^{*}_{\rm EC}$. 

To decide the nature of the fixed point $\vec{x}^{*}_{\rm EC}$, we focus on the area where $b \gtrsim 1.0$, as depicted in the inset of Fig. \ref{fig:Trajectories}.
%, to determine the nature of the fixed point $\vec{x}^{*}_{\rm EC}$.
Observations of this area suggest that the system experiences a ``random force," causing the trajectory to resemble a two-dimensional random walk. 
We believe that the origin of this random term might be associated with the number of realizations of disorder considered here. As noted in Ref. \cite{Oganesyan:2007}, the number of realizations of the disorder required to achieve statistical confidence varies significantly with the strength of the disorder.
The interaction between this random element and the deterministic flow pattern anticipated with an increase in the number of realizations can be modeled by the linearized system of SDEs

\begin{equation}
    \mathrm{d}\vec{X} = \boldsymbol{\mathcal{A}} \vec{X} \mathrm{d}t + R \mathrm{d} \vec{\Omega},
    \label{eq:Lin-SDE}
\end{equation}
where $\vec{X} \equiv \left(\beta - \beta^*, \gamma - \gamma^*\right)^\top$, the deterministic part is described by the following matrix
\begin{equation}
    \boldsymbol{\mathcal{A}} = 
    \begin{pmatrix}
        a & \eta - q  -\delta \\
        \frac{1}{\eta + q  +\delta} & a-2\delta
    \end{pmatrix},
    \label{deterministic-matrix}
\end{equation}
%in terms of the parameters $a$,  $\eta$ $q$ and $\delta$. 
%For 
where $a < -1$, $\eta=1$ and $q=\delta=0$, the fixed point is a stable node. 
For $a < 0$, $\eta=0$, with $q=1$ and $\delta=0$ the fixed point behaves as a stable spiral, while $a < 0$, $\eta=0$, $q>1$ and $\delta=1$ it is a degenerate node. $\vec{\Omega} \equiv \left(\Omega_{\beta},\Omega_{\gamma}\right)^\top$, where
 $\Omega_{\beta}$ and $\Omega_{\gamma}$ are independent Wiener processes. The derivative of a Wiener process represents a white noise. The coupling of these random processes with the deterministic parts of the equations is governed by the parameter $R$, which modulates the intensity of the noise. For $R=0$, the system of SDEs becomes a linearized system of coupled ordinary differential equations (ODEs). 
 %(ODEs).

We fit the data (red triangles) presented in the inset of Fig. \ref{fig:Trajectories} 
%with the system of ODEs ($R=0$) described in Eq. (\ref{eq:Lin-SDE}). 
using Eq. (\ref{eq:Lin-SDE}) with $R=0$.
%with the {\tt lmfit} package. 
The result suggests that $\vec{x}^{*}_{\rm EC}$ is a degenerate node with best-fit parameters $a=-1.7(2)$, $q=15(2)$, $\beta^{*} = 0.695(6)$ and $\gamma^{*} = 1.186(3)$. This result is depicted as a solid black curve in the inset.
%of Fig. \ref{fig:Trajectories}. 
We also present ten random trajectories (gray curves) obtained from the system of SDEs with the best-fit parameters and $R=0.02$. We observe that these random trajectories are qualitatively similar to the 
trajectory obtained from the fitting procedure of our empirical data (red curve).
%red one 
%obtained by fitting the averaged distributions with the surmise expression Eq. (\ref{General_ratio_dist}). 

Our result suggests the emergence of a non-ergodic phase distinct from the MBL as $b \gg 1$ characterized by the fixed point $\vec{x}^{*}_{\rm EC}$. 
%This aligns with the results in Ref. \cite{Siegl:NJP2023} which utilized the sample-to-sample variance of the 
%expected value of the ratio of consecutive level spacing 
%averaged consecutive-gap ratio $\braket{r}$
%to identify this phase as an incomplete many-body localized phase. 
In contrast to the local field case, the exchange coupling case does not attain the MBL phase due to the presence of the spin rotation symmetry, indicating that $\beta$ neither approaches zero nor $\gamma$ tends to one regardless of the magnitude of $b$. 
%In the following, we discuss the stability of the fixed point $\vec{x}^{*}_{\rm EC}$ 

\subsection{Stationary probability distribution stability in a continuous-state Markov process}

%\textcolor{red}{PUT A MOTIVATION PARAGRAPH HERE! WHAT WE ARE GOING TO SHOW HERE?}

For stochastic systems, the idea of stability of fixed point does not make sense, so one has to study the evolution of the probability distribution of finding the system close to the fixed point as time goes on. 
This can be realized by considering the system of SDEs described by Eq. (\ref{eq:Lin-SDE}) as a {\it continuous-state Markov process} \cite{stochastic-dynamics:book1994},
governed by the discrete-time dynamics of the probability distribution given by the following master equation
\begin{equation}
    P(\vec{X}_{k+1}) = \int \mathcal{P}(\vec{X}_{k+1}\mid\vec{X}_{k}) P(\vec{X}_{k}) \mathrm{d}\vec{X}_{k},
    \label{eq:master_equation}
\end{equation}
where $\vec{X}_{k} \equiv \vec{X} (t= k \Delta t)$ and $\Delta t$ is a small increment of time. $P(\vec{X}_{k})$ is the probability of finding the particle in $\vec{X}_k$. $\mathcal{P}(\vec{X}_{k+1}\mid\vec{X}_{k})$ is the conditional probability of finding the particle in $\vec{X}_{k+1}$ given that it was in $\vec{X}_{k}$.

%Our aim here is to find 
We look for a stationary distribution that is the fixed point of this master equation.
Let us assume that the initial probability is given by a bivariate Gaussian distribution with the mean value $\vec{\mu}_0 = \vec{X}^* \equiv (0,0)^\top$ and the covariance matrix $\boldsymbol{\Sigma}_0$, $\vec{X}_0 \sim \mathcal{N}(\vec{0},\boldsymbol{\Sigma}_0)$. We assume that conditional probability $\mathcal{P}(\vec{X}_{k+1}\mid\vec{X}_{k})$ can also be written as a bivariate Gaussian distribution 
with  
%\textcolor{blue}{characterized by the}
mean value $\vec{\mu}_{k+1} = \boldsymbol{A} \vec{X}_k$, 
%\textcolor{blue}{$\vec{\mu}_{k+1} = \boldsymbol{A} \vec{X}_k+R\vec{\omega}_k$}, 
where $\boldsymbol{A} \equiv \boldsymbol{1} + \boldsymbol{\mathcal{A}} \Delta t$ %\textcolor{blue}{and $\vec{\omega}_k$ the white noise increment, with}
and covariance matrix 
%\textcolor{blue}{given by}
$\boldsymbol{\Sigma}$, 
$\vec{X}_{k+1}\mid\vec{X}_k \sim \mathcal{N} (\vec{\mu}_{k+1},\boldsymbol{\Sigma})$.

Since the integral of the product of two multivariate Gaussian distributions is also a multivariate Gaussian distribution, $P(\vec{X}_1)$ obtained by Eq. (\ref{eq:master_equation}) is also a bivariate Gaussian distribution given by $\vec{X}_1 \sim \mathcal{N}(\vec{0},\boldsymbol{\Sigma}_1)$, where $\boldsymbol{\Sigma}_1 = \boldsymbol{\Sigma} + \boldsymbol{A}^\top \boldsymbol{\Sigma}_0 \boldsymbol{A}.$ 
Therefore, the solution 
%for the $k$-th step, 
$P(\vec{X}_{k})$ is also a bivariate Gaussian distribution $\vec{X}_k \sim \mathcal{N}(\vec{0},\boldsymbol{\Sigma}_k)$,
where its covariance matrix can be determined by recursively solving the matrix equation system 
\begin{equation}
\boldsymbol{\Sigma}_{k+1} = \boldsymbol{\Sigma} + \boldsymbol{A}^\top \boldsymbol{\Sigma}_k \boldsymbol{A}.
\label{eq:discrete-system}
\end{equation}
%\textcolor{blue}{
%If $\boldsymbol{\Sigma}_0$ is the fixed point of the discrete system Eq. (\ref{eq:discrete-system}) $\boldsymbol{\Sigma}^*$, then $\boldsymbol{\Sigma}_k=\boldsymbol{\Sigma}^*$.
%IMPROVE/SIMPLIFY THIS PART!}
The fixed point of this discrete system, $\boldsymbol{\Sigma}_{k+1}=\boldsymbol{\Sigma}_k=\boldsymbol{\Sigma}^*$, is obtained by solving the discrete Lyapunov equation 
\begin{equation}
    \boldsymbol{\Sigma}^* = \boldsymbol{\Sigma} + \boldsymbol{A}^\top \boldsymbol{\Sigma}^* \boldsymbol{A}.
    \label{discrete_lyapunov_equation}
\end{equation}

\begin{figure}[htbp]
    \centering
    \includegraphics[width=0.5\textwidth]{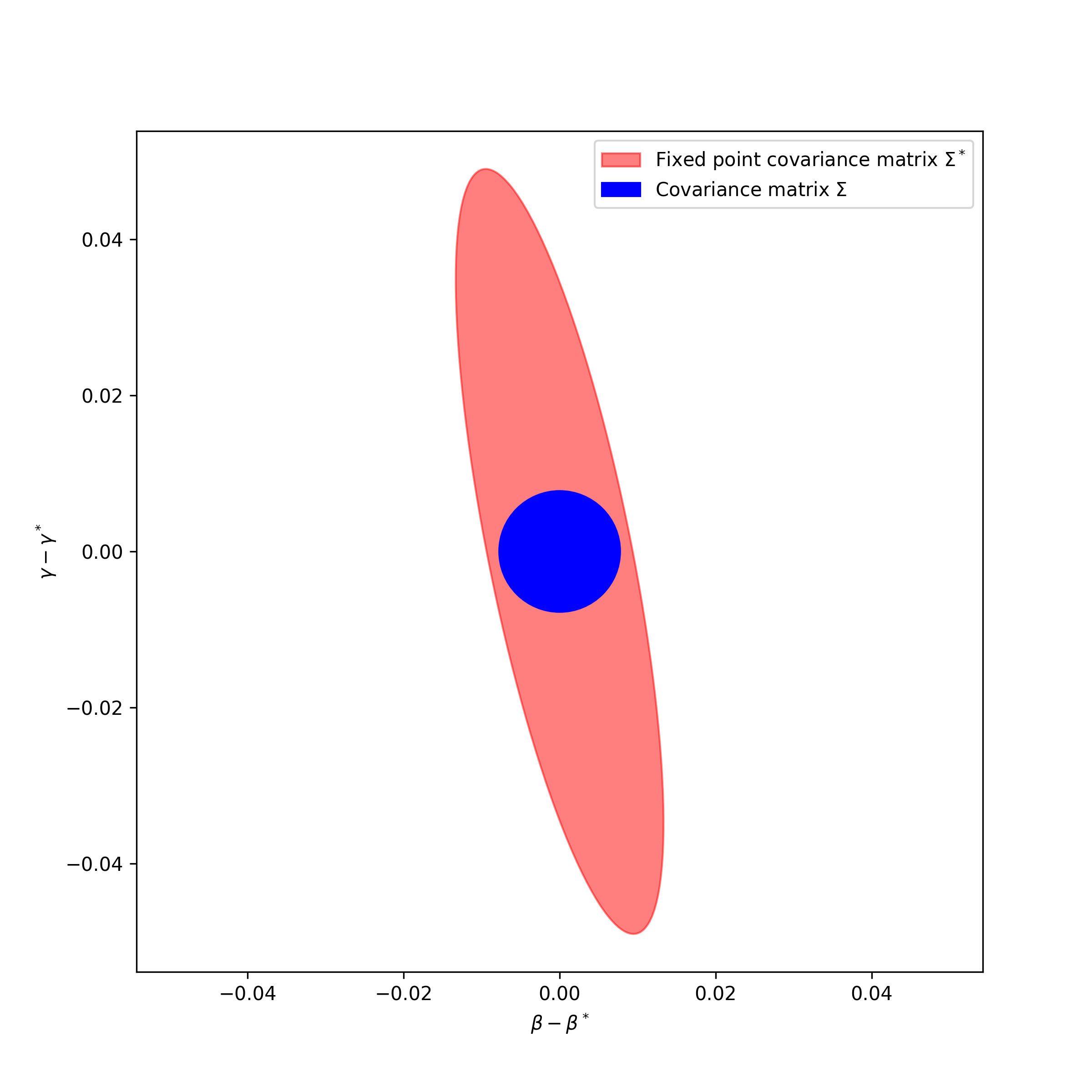}
    \caption{The red ellipse delineates the $95\%$ confidence region corresponding to the stationary probability distribution $P(\vec{X})$. This ellipse is centered at the fixed point $\vec{x}^{*}_{\rm EC} = (\beta^*,\gamma^*)^\top \approx (0.7,1.2)^\top$ and characterized by the covariance matrix $\boldsymbol{\Sigma}^*$ as specified in Eq. (\ref{Sigma_star}). In contrast, the blue ellipse depicts the $95\%$ confidence region for a distribution with  $\boldsymbol{\Sigma}={\rm diag}(10^{-5},10^{-5})$.
    }
    \label{Confidence_ellipses}
\end{figure}

To verify the stability of the fixed point $\vec{x}^{*}_{\rm EC}$, we consider the deterministic matrix $\boldsymbol{\mathcal{A}}$, Eq. (\ref{deterministic-matrix}). This matrix is evaluated with the best-fit parameters obtained previously, $\Delta t =0.1$ and a positive definite matrix $\boldsymbol{\Sigma}={\rm diag}(\sigma^2,\sigma^2)$. We use $\sigma^2=10^{-5}$ estimated from the standard errors of the $\beta$ and $\gamma$ parameters obtained from the fitting procedure.
From Eq. (\ref{discrete_lyapunov_equation}), we obtain the following positive definite covariance matrix 
\begin{equation}
    \boldsymbol{\Sigma^*} = 
    \begin{pmatrix}
        3\times 10^{-5} & - 8\times 10^{-5} \\
        - 8\times 10^{-5} & 4\times 10^{-4}
    \end{pmatrix}.
    \label{Sigma_star}
\end{equation}

Figure \ref{Confidence_ellipses} illustrates the $95\%$ confidence ellipses pertaining to the bivariate normal distribution, which is centered at the fixed point $\vec{x}^{*}_{\rm EC}$. The ellipse depicted in red corresponds to $\boldsymbol{\Sigma}^*$, representing the stationary probability distribution $P(\vec{X}_{k})$. In contrast, the ellipse illustrated in blue denotes $\boldsymbol{\Sigma}$, which is the covariance matrix associated with the conditional probability distribution $\mathcal{P}(\vec{X}_{k+1}\mid\vec{X}_{k})$.

As anticipated, stochastic fluctuations tend to disperse the probability distribution $P(\vec{X}_{k})$ over the $\beta\gamma$ parameter space. However, the deterministic component, represented by the matrix $\mathcal{A}$ and evaluated using the best-fit parameters obtained from our numerical data, effectively maintains this probability distribution concentrated in the vicinity of the fixed point.
Thus, the probability of finding the system close to the fixed point $\vec{x}^{*}_{\rm EC}$ as the maximal exchange coupling strength $b$ increases is given by the bivariate normal distribution $\vec{X} \sim \mathcal{N}(\vec{0},\boldsymbol{\Sigma}^*)$.

\section{Summary and Outlook} \label{sec_conclusion}

In this paper, we proposed a two-parameter surmise expression 
%Eq. (\ref{General_ratio_dist}) 
for the 
%ratio of consecutive level spacing 
consecutive-gap ratio
distribution of many-body Hamiltonians
undergoing a GOE-to-Poisson statistics crossover. 
%Inspired by the surmise expression for the 
%ratio of consecutive level spacing 
%\textcolor{blue}{consecutive-gap ratio}
%distribution of the Wigner-Dyson ensembles \cite{Atas:PRL2013} and the crossover expression for the level spacing distribution proposed in Ref. \cite{Serbyn:PRB2016}, our two-parameter distribution effectively captures the crossover between GOE and Poisson statistics. 
We applied this expression to study the MBL transition in an isotropic Heisenberg spin-$1/2$ chains with disordered local field and exchange couplings. Our main findings demonstrate the efficacy of our proposed distribution in describing the MBL transition, providing valuable insights into the spectral properties of complex %quantum 
systems.

%We compare Eq. (\ref{General_ratio_dist}) with numerical data from the ensemble-averaged consecutive-gap ratios for both the local field and exchange coupling cases, showing excellent agreement.\textcolor{red}{
%Using a classification scheme similar to that in Ref. \cite{Serbyn:PRB2016}, we describe the GOE-to-Poisson statistics crossover in three stages based on the parameters $\beta$ and $\gamma$. The local field case traverses all three stages as the maximal local field strength $h$ increases, culminating in the 
%MBL phase, while the exchange coupling case does not reach the third stage, indicating the absence of full localization, even for a high maximum exchange coupling strength $b$. The concept of stages provides a unified framework for classifying the intermediate behaviors of different models as they transition from an ergodic to a localized phase.}

%This study reveals key differences between the disorder models, indicated by the trajectories of the best-fit parameters in the $\beta\gamma$ space shown in Fig. \ref{fig:Trajectories}.
Drawing upon an analogy with nonlinear dynamics, we 
%regard $h$ and $b$ as representations of time, and $\beta$ and $\gamma$ as phase-space coordinates, to 
examined the GOE-to-Poisson statistics in disordered systems through the lens of dynamical flows. This framework offers a compelling interpretation of the statistical evolution of disordered quantum systems, linking their spectral properties to established notions in nonlinear dynamics.
For the local field case, the flow pattern converges to the fixed point %$\vec{x}^{*}_{\rm LF} = (0,1)^\top$,
corresponding to the MBL phase. 
%characterized by Poisson statistics. 

In contrast, the exchange coupling case exhibits a distinct fixed point 
%at $\vec{x}^{*}_{\rm EC} \approx (0.7,1.2)^\top$, 
associated with a non-ergodic phase different from the MBL phase due to the %$SU(2)$ 
spin rotation symmetry present in this system. 
We introduce a system of SDEs
%, Eq. (\ref{eq:Lin-SDE}), 
to describe the behavior of the system near the fixed point. 
%$\vec{x}^{*}_{\rm EC}$.
Within the stochastic regime, the notion of stability of fixed points is replaced by the stationary probability distribution of 
%finding the system close to the fixed point as the dynamics progress.
a continuous-state Markov process.
%as time goes on, which for large times ($k \gg 1$) 
%for large times.
%approaches a bivariate normal distribution $\mathcal{N}(\vec{0},\boldsymbol{\Sigma}^*)$, which is a fixed point of a Markov process described by the master equation Eq. (\ref{eq:master_equation}). 
We demonstrate that, under the assumption of a normally distributed transition probability distribution, the resulting stationary probability distribution converges to a bivariate normal distribution.
%\textcolor{red}{CONSIDER TO INCLUDE AN IMPROVED VERSION OF THE COMMENTED SENTENCE BELOW!}
For further works, we are interested in investigating whether other forms of transition probability distributions may affect the stationary probability distribution.
%approach to understand
%of the interplay between random forces and deterministic flows in the system.

\begin{comment}

Finally, we employ a finite-size scaling technique to estimate the critical disorder $h_{\rm c}$ where the MBL transition occurs for the local field case. 
We use the best fit parameter $\beta$ of our surmise expression for various chain lengths to proceed with the finite-size scaling analysis. We estimate a critical disorder strength $h_{\rm c} = 2.8(5)$, which is consistent with previous estimates %($h_{\rm c} \approx 2.6,\, ...,\, 3.0$) 
obtained 
%using the sample-to-sample variance statistics of the averaged 
%ratio of consecutive level spacing, 
%\textcolor{blue}{consecutive-gap ratio},
%as reported 
in Refs. \cite{Schliemann:PRB2021,DeLuca_Scardicchio:EPL2013}.

\end{comment}

This probabilistic framework might be useful to elucidate the evolution of spectral properties in many-body complex systems experiencing a crossover as a parameter, such as disorder strength, in the current study.
%We believe that the surmise distribution proposed here can be useful not only for studying the spectral properties of quantum many-body systems but also for studying real-world complex networks. 
%Over the years, 
The spectral properties of real-world complex networks have been extensively studied to gain insight into the underlying mechanisms governing these systems. 
%For example, 
Researchers have examined protein-protein interaction networks to gain insight into biological processes \cite{aguiar-baryam:pre2005,Bandyopadhyay-Jalan:PRE2007}, and the study of spectral correlations in the price variations of the financial market has provided a valuable understanding of market dynamics \cite{laloux_etal:prl1999,plerou_etal:prl1999,plerou_etal:pre2002}. 
By applying our surmise and our stochastic dynamical system approach to characterize crossovers, we can potentially uncover new patterns and behaviors in these diverse fields. 
Our proposed methodology thus holds promise for advancing the understanding of spectral properties in various complex real-world networks.

\section*{Acknowledgments}

This work was supported by the São Paulo Research Foundation (FAPESP) Grant No. 2020/00841-9, and from Conselho Nacional de Pesquisas (CNPq), Grant No. 301595/2022-4. 
GCDF acknowledges Professors José Carlos Egues and Miled Moussa for all the support provided during his one-year period at the Instituto de Física de São Carlos at Universidade de São Paulo (IFSC). GCDF also thanks the staff of the High Performance Cluster (HPC) at Universidade de São Paulo (USP) for their support during this time. 
%\textcolor{red}{Grant numbers? Other acknowledgments? Julian's new affiliation?}

%\bibliographystyle{unsrt}

%References 
%\bibliography{sample.bib}% Produces the bibliography via BibTeX.

%\begin{comment}

\bibliographystyle{unsrt}

%\begin{comment}

%\end{comment}

\end{document}